\documentclass[
reprint,
superscriptaddress,
frontmatterverbose,
preprintnumbers,
nofootinbib,
nobibnotes,
amsmath,amssymb,
aps,prc
]{revtex4-1}
\usepackage{graphicx}
\usepackage{times}
\usepackage{braket}
\usepackage{physics}
\usepackage{CJK}

\def\fun#1#2{\lower3.6pt\vbox{\baselineskip0pt\lineskip.9pt
 \ialign{$\mathsurround=0pt#1\hfil##\hfil$\crcr#2\crcr\sim\crcr}}}
\usepackage{dcolumn}
\usepackage{bm}

\usepackage{ulem}
\usepackage{color}
\definecolor{verde}{rgb}{0.0,0.7,0.0}

\begin{document}
\begin{CJK}{UTF8}{ipxm}


\title{Calculation of $\beta$-decay half-lives with Skyrme Hartree-Fock-Bogoliubov+$pn$-QRPA and isoscalar pairing strengths optimized by a Bayesian method}


\author{Futoshi Minato (湊 太志)}
\email{minato.futoshi@jaea.go.jp}
\affiliation{Nuclear Data Center, Japan Atomic Energy Agency, Tokai, Ibaraki 319-1195, Japan}

\author{Z. M. Niu (牛中明)}
\email{zmniu@ahu.edu.cn}
\affiliation{School of Physics and Material Science, Anhui University, Hefei 230601, China}
\affiliation{Institute of Physical Science and Information Technology, Anhui University, Hefei 230601, China}

\author{Haozhao Liang (梁豪兆)}
\email{haozhao.liang@phys.s.u-tokyo.ac.jp}
\affiliation{Department of Physics, Graduate School of Science, The University of Tokyo, Tokyo 113-0033, Japan}
\affiliation{RIKEN iTHEMS, Wako 351-0198, Japan}

\date{\today}

\begin{abstract}
 \noindent
 \textbf{Background:} 
For radioactive nuclear data, $\beta$ decay is one of the most important information and is applied to various fields.
However, some of the $\beta$-decay data are not available due to experimental difficulties. 
From this respect, theoretically calculated results have been embedded in the $\beta$-decay data to compensate the missing information.
\\
 \textbf{Purpose:} 
Theoretical $\beta$-decay calculations are required to treat various nuclear correlations as precise as possible. 
In particular, the pairing correlation is one of the most important factors to reproduce the $\beta$-decay half-lives correctly. 
First of all, we study the effect of zero- and finite-range isovector pairings on half-lives.
We then study the isoscalar pairing strengths, which are determined through experimental data of half-life, and finally predict the isoscalar pairing strengths and half-lives of neutron-rich nuclei.
\\
 \textbf{Methods:} 
To calculate the $\beta$-decay half-lives, a proton-neutron quasi-particle random phase approximation on top of a Skryme energy density functional is applied with an assumption of spherical symmetry.
The half-lives are calculated by including the allowed and first-forbidden transitions.
The isoscalar pairing strength is estimated by a Bayesian neutral network (BNN).
We verify the predicted isoscalar pairing strengths by preparing the training data and test data.
\\
 \textbf{Results:}
It was confirmed that the finite-range isovector pairing ensures the $\beta$-decay half-lives insensitive to the model space, while the zero-range one was largely dependent on it.
The half-lives calculated with the BNN isoscalar pairing strengths reproduced most of experimental data, although those of highly deformed nuclei were underestimated.
We also studied that the predictive performance on new experimental data that were not used for the BNN training and found that they were reproduced well.
\\
\textbf{Conclusions:} 
Our study demonstrates that the isoscalar pairing strengths determined by the BNN can reproduce experimental data in the same accuracy as other theoretical works.
To achieve a more precise prediction, the nuclear deformation is important.

\end{abstract}

\maketitle

\end{CJK}

\section{Introduction}
\label{sect:Intro}
$\beta$ decay, the representative decay mode of unstable nuclei, was first recognized at the end of 19th century.
The discovery greatly extended the field of nuclear chemistry and led to opening nuclear physics.
Nowadays this phenomenon becomes more important for various fields beyond nuclear chemistry and physics, such as radiology, geoscience, nuclear engineering, and astrophysics.
Needless to say, $\beta$ decay has attracted a lot of attentions of researchers since its discovery.
In the past decades, the study of $\beta$ decay has been extended into very neutron-rich nuclei with interests in the exotic nuclear structure~\cite{Otsuka2020} and for a finer understanding of the $r$-process nucleosynthesis~\cite{Kajino2017}, which is a promising scenario synthesizing elements heavier than iron in star evolutions.
\par
Recent progress on experiments for unstable nuclei has accelerated our understanding of $\beta$ decay and provides the high-accuracy nuclear decay data.
Several measurements of half-lives ($T$) and $\beta$-delayed neutron emission branching ratios start to cover the nuclei relevant to the $r$ process~\cite{Lorusso2015, RIKEN2020}.
However, there still remain a lot of unmeasured neutron-rich unstable nuclei important for the $r$ process.
In particular, the $\beta$-decay data on nuclei locating at the south-east region from $^{208}$Pb in the nuclear chart and on neutron-rich actinides that are fissionable are severely absent.
To compensate unmeasured data, it is necessary to use nuclear theoretical approaches for the $r$-process simulation.
\par
Because a nucleus is a finite many-body system composed of nucleons, the calculation requires more or less model approximations.
To predict the $\beta$-decay half-lives as accurate as possible, building a model with less phenomenological treatment is important.
One of the candidates suitable for this object is a microscopic model based on a two-body effective interaction.
In particular, a self-consistent framework, which means the use of same interaction between the ground and excited states, is considered to be the reasonable approach reducing ambiguities on the interactions.
A lot of studies within this framework have been carried out to study the $\beta$-decay half-lives for specific isotopic or isotonic chains, {\it e.g.}, the proton-neutron quasiparticle random-phase approximation ($pn$QRPA) or finite amplitude method ($pn$FAM) on the top of the energy-density functional (EDF) ~\cite{Engel1999, Niksic2005, Yoshida2013, Niu2013, PhysRevC.89.064320, Martini2014, Sarriguren2015, Minato2016, Wang2016, Mustonen2016, Borzov2020, Wen2021}, the configuration interaction methods~\cite{Suzuki2012, Suzuki2018, Suzuki2019}, and the interacting boson model~\cite{Nomura2020}.
Among them, from the viewpoint of computational feasibility, the $pn$QRPA and $pn$FAM are currently the only methods that have been applied to a systematical prediction of $\beta$-decay half-lives including the first-forbidden (FF) transitions over the nuclear chart~\cite{Tomislav2016,Ney2020}.
\par
To predict the $\beta$-decay half-lives, one needs a careful attention to the pairing correlation, which accounts for a short-range attractive interaction between nucleons that is not taken into account in the Hartree-Fock level.
The contributions from the pairing correlation to the half-lives are threefold: (1) additional binding energies to nuclei, (2) variation of particle occupation probabilities, and (3) supplement of particle-particle residual interaction.
For the $pn$QRPA on the top of the Skyrme EDF, the zero-range interactions have conventionally been used as the isovector spin-singlet and isoscalar spin-triplet pairing forces due to their simpleness.
However, it was pointed out that the half-lives calculated with an isoscalar zero-range interaction is largely different depending on a model space through (3), and moreover the deviations become larger with increasing strength of isoscalar pairing interaction~\cite{Engel1999}.
Other studies~\cite{Wang2016, Minato2016, Tomislav2016} also indicated that different strengths of the isoscalar pairing interaction are needed to reproduce experimental data of different isotopic or isotonic chains.
These results pose a question on the effectiveness of the zero-range isoscalar pairing interaction for the reliable predictions of half-lives.
The same question also arises in the isovector pairing that mainly affects the $\beta$ decay through the aforementioned (1) and (2).
\par
The isoscalar pairing has an effect of considerably reducing the excitation energies of the low-lying Gamow-Teller (GT) transitions.
Since the $\beta$ decay for nuclei from light to heavy mass ($Z\approx82$) is largely invoked by the GT transitions, this reduction enlarges the energy released by the $\beta$ decay, that is the so-called end-point energy, reducing the half-lives significantly~\cite{Engel1999, Niksic2005, Marketin2007, Yoshida2013, Wang2016}.
In the half-life calculation, the isoscalar pairing strength has been treated as a free parameter independent of the effective force used in the ground-state calculation, and has been adjusted so as to reproduce the experimental half-lives.
We have an interest in how the isoscalar pairing strength evolves with increasing proton and neutron numbers, and try to estimate it for neutron-rich nuclei.
A better description for the pairing interactions is the use of finite-range force.
This force naturally includes a cutoff in the pairing model space~\cite{Takahara1994} avoiding the so-called ultraviolet divergence and yields half-lives less sensitive to the model space~\cite{Engel1999}.
The finite-range pairing forces have been already applied to study the $\beta$ decay within the covariant density functional (CDF)+$pn$ relativistic QRPA ($pn$RQRPA)~\cite{Tomislav2016} and the Gogny EDF+$pn$QRPA~\cite{Martini2014}, while its application to the Skyrme Hartree-Fock-Bogoliubov (HFB)+$pn$QRPA is limited to some isotopes and isotones, and only for the isoscalar channel~\cite{Engel1999}.
\par
The purpose of this work is to study the isoscalar pairing strengths of neutron-rich nuclei for the systematical prediction of $\beta$-decay half-lives. 
To this end, we construct a Skyrme HFB+$pn$QRPA with a finite-range pairing force to reduce the uncertainties coming from the pairing correlations.
We determine the isoscalar pairing strengths so as to reproduce the experimental half-lives, and estimate those of neutron-rich nuclei with no experimental data.
In particular, we apply a Bayesian neural network (BNN), which has been applied to predict nuclear masses~\cite{Utama2016Phys.Rev.C93_014311, ZM2018} and $\beta$-decay half-lives~\cite{ZM2019}, for the estimation of isoscalar pairing strengths.
We assess the performance of the strengths calculated by the BNN and discuss the result quantitatively.
It should be mentioned that this work corresponds to a non-relativistic counterpart of the CDF+$pn$RQRPA~\cite{Tomislav2016} although they use the isospin-dependent force~\cite{Niu2013} of the isoscalar pairing.
\par
This paper is organized as follows. 
In Sect.~\ref{sect:theory}, we describe the theoretical framework to calculate the $\beta$-decay half-lives using the Skyrme HFB+$pn$QRPA. 
In Sect.~\ref{sect:result}, the results obtained in this work are presented and discussed comparing with the experimental data and preceding works.
Section~\ref{sect:summary} summarizes this work and presents some perspectives.
The complete data table containing the calculated half-lives is available in Supplemental Material.
%
\section{Theoretical Framework}
\label{sect:theory}
%
\subsection{Skyrme HFB+$pn$QRPA with a finite-range pairing force}
\label{sect:pnQRPA}
The $\beta$-decay calculations of the Skyrme HFB+$pn$QRPA are separated into two parts: the ground and excited states.
We begin with calculating the ground state of nuclei, that is equal to an energy minimum against small-amplitude surface vibrations, within the HFB approach.
In this work, we use SkO'~\cite{Reinhard1999} for the effective particle-hole two-body interaction, which is known as giving a reasonable agreement with the experimental $Q_{\beta}$ values~\cite{Mustonen2016}.
The pairing correlations are treated by considering the finite-range effect, and the Gogny-type interaction,
\begin{equation}
V_{pp}^{(1)}(\vb*{r}_1,\vb*{r}_2)=\sum_{i=1}^2\Big(W_{i}+B_{i}P_{\sigma}-H_{i}P_{\tau}-M_{i}P_{\sigma} P_{\tau}\Big)e^{-r_{12}^{2}/\mu_{i}^{2}},
\label{eq:finiteforce}
\end{equation}
is used for the isovector particle-particle channel, where $r_{12} = |\vb*{r}_{1} - \vb*{r}_{2}|$, and $P_{\sigma}$ and $P_{\tau}$ are the spin and isospin exchange operators, respectively.
The parameters $W_{i}$, $B_{i}$, $H_{i}$, $M_{i}$, and $\mu_{i}$ are taken from the D1S force~\cite{Berger1984}.
An advantage of using the finite-range force, {\it e.g.}, D1S, is that it automatically introduces a natural cutoff in the momentum space for the particle-particle scattering and is capable of avoiding an ultraviolet divergence, which occurs in the zero-range forces~\cite{Garrido1999, Satula2006}.

The quasiparticle states, denoted by $k=\{n,j,l\}$, where $n$ is the principal quantum number, and $j$ and $l$ are the total and orbital angular momenta, respectively, are obtained by solving the HFB equation
\begin{equation}
\begin{split}
\int d\vb*{r}_{2}&
    \left(
    \begin{array}{cc}
         h_{q}(\vb*{r}_{1},\vb*{r}_{2})-\lambda_{q} & \Delta_{q}(\vb*{r}_{1},\vb*{r}_{2})\\
        \Delta_{q}(\vb*{r}_{1},\vb*{r}_{2}) & -h_{q}(\vb*{r}_{1},\vb*{r}_{2})+\lambda_{q}
    \end{array}
    \right)
    \left(
    \begin{array}{c}
        U_{k}(\vb*{r}_{2}) \\
        V_{k}(\vb*{r}_{2})
    \end{array}
    \right)\\
    &=
    E_{k}
    \left(
    \begin{array}{c}
        U_{k}(\vb*{r}_{1})\\
        V_{k}(\vb*{r}_{1})
    \end{array}
    \right),
\end{split}
\label{eq:HFB}
\end{equation}
where $h$ and $\Delta$ are calculated by the first derivatives of energy functional with respect to the normal and pairing densities, respectively~\cite{Dobaczewski1984}, and $\lambda_{q=n,p}$ are the nucleon Fermi energies.
We expand $U$ and $V$ as
\begin{equation}
\begin{split}
U_{k}(\vb*{r})&=\sum_{l} U_{lk}\,\varphi_{l}(\vb*{r}),\\ V_{k}(\vb*{r})&=\sum_{l} V_{lk}\,\varphi_{\bar{l}}(\vb*{r}),    
\end{split}
\label{eq:UV}
\end{equation}
where the basis functions $\{\varphi_k\}$ are obtained by solving $\int d\vb*{r}'\,h(\vb*{r},\vb*{r}')\varphi_{k}(\vb*{r}')=\varepsilon_{k}\varphi_{k}(\vb*{r})$ in the coordinate space in order to properly describe the asymptotic behavior of densities~\cite{Stoitsov2003}. 
Note that our calculation is carried out assuming the spherical symmetry.
For the practical calculation of Eq.~\eqref{eq:HFB}, we truncate the expansion of $U$ and $V$ at a point where $\varepsilon_{k}$ is smaller than a cutoff energy $\varepsilon_{\rm cut}$.
The pairing potential is defined as $\Delta(\vb*{r}_{1},\vb*{r}_{2})=\sum_{k'l'}V_{pp}^{(1)}(\vb*{r}_{1},\vb*{r}_{2}) \kappa_{k'l'}(\vb*{r}_{1},\vb*{r}_{2})$, where $\kappa$ is the pairing density.
The continuum states are discretized by the radial box of $20$~fm with a step size being $\Delta r=0.1$~fm.
\par
The excited states of daughter nuclei resulted from $\beta$ decay are calculated with the $pn$QRPA in the canonical basis of the HFB.
The $pn$QRPA equation is given by the following eigenvalue problem:
\begin{equation}
    \sum_{p'n'}\left(
    \begin{tabular}{rr}
    $A_{pnp'n'}^{(c)}$ & $B_{pnp'n'}^{(c)}$\\
    $-B_{pnp'n'}^{(c)*}$ & $-A_{pnp'n'}^{(c)*}$
    \end{tabular}
    \right)
    \left(
    \begin{tabular}{c}
    $X^{(c)}_{p'n'}$\\
    $Y^{(c)}_{p'n'}$
    \end{tabular}
    \right)=
    \mathcal{E}_{c}
    \left(
    \begin{tabular}{c}
    $X^{(c)}_{pn}$\\
    $Y^{(c)}_{pn}$
    \end{tabular}
    \right).
    \label{eq:QRPA}
\end{equation}
Here, $A^{(c)}$ and $B^{(c)}$ are matrix elements including the particle-hole and particle-particle interactions given in the canonical basis~\cite{Engel1988, Engel1999}, and the subscript $c=(1^{+}, 0^{-}, 1^{-}, 2^{-})$ represents the $\beta$-decay type of the allowed and FF transitions.
The eigenvalues $\mathcal{E}_{c}$ are used for calculating the excitation energies of daughter nucleus and the eigenvectors $X^{(c)}$ and $Y^{(c)}$, which respectively correspond to the forward and backward amplitudes of $pn$QRPA, are used to calculate the transition strengths of $\beta$ decay.
\par
For the isoscalar particle-particle residual interaction, we use the two-Gaussian force~\cite{Engel1999,Tomislav2016}
\begin{equation}
V_{pp}^{(0)}(\vb*{r}_{1},\vb*{r}_{2})=-V\sum_{i=1,2}g_{i} \exp\left(-\frac{r_{12}^{2}}{\mu_{i}'^{2}}\right)\hat{\Pi}_{S=1,T=0},
\label{eq:isp}
\end{equation}
where $\hat{\Pi}_{S=1, T=0}$ is the projection operator on the isoscalar spin-triplet channel, and $g_{1}=1$, $g_{2}=-2$, $\mu_{1}'=1.2$~fm, and $\mu_{2}'=0.7$~fm, which are chosen so that $V_{pp}^{(0)}$ is repulsive at small distance and attractive at long distance. 
The parameter $V$ is the isoscalar pairing strength.
\par
To carry out the diagonalization of Eq.~\eqref{eq:QRPA}, we consider the single-particle energy in the canonical basis up to $\varepsilon_{\rm cut}=40$~MeV and the two-quasiparticle energy up to $E_{p}+E_{n}=80$~MeV, which is enough to obtain a stable result in terms of model space as described later. 
\par
For calculating the odd-mass nuclei, the same formalism of the even-even nuclei is applied, as adopted in Ref.~\cite{Tomislav2016}, namely the average particle number is adjusted so as to reproduce the number of nucleons in interest by tuning $\lambda_{q}$ in Eq.~\eqref{eq:HFB}.
We should mention that a better treatment of the odd-mass nuclei can be achieved by using the equal filling approximation as like Ref.~\cite{Ney2020}, and this is the plan for our next works.
\subsection{$\beta$-decay half-life}
\label{sect:half}
The $\beta$-decay rate to a daughter nucleus state, denoted by $\gamma$, is calculated by~\cite{Behrens}
\begin{equation}
  \lambda_{\beta}^{(c,\gamma)}
  =\frac{\ln 2}{T^{(c,\gamma)}}\\
  =\frac{\ln 2}{D}\int_0^{p_{0}}
  p_e^2(W_0-W)^2F(Z,W)C(W)dp_e,
  \label{eq:rate}
\end{equation}
where $W=\sqrt{(p_ec)^2+1}$ and $p_ec$ are the electron energy in terms of the electron mass unit and the electron momentum in terms of $m_ec$, respectively.
The physical constant $D=6144.4\pm2.0$~s is taken from Ref.~\cite{Hardy2009}. 
The maximum electron energy is defined as $W_0=(Q_{\beta}-\mathcal{E}_{c,\gamma}^*)/(m_e c^2)$, where the numerator $Q_{\beta}-\mathcal{E}_{c,\gamma}^*$ is called the end-point energy.
$Q_{\beta}$ and excitation energy of the daughter nucleus $\mathcal{E}_{c,\gamma}^*$ are approximated as~\cite{Engel1999}
\begin{equation}
    Q_{\beta}=\lambda_{n}-\lambda_{p}+\Delta M_{n-H}-E_{\rm corr}
    \label{eq:Q}
\end{equation}
and
\begin{equation}
    \mathcal{E}_{c,\lambda}^*=\mathcal{E}_{c,\lambda}-E_{\rm corr},
    \label{eq:Ex}
\end{equation}
respectively, and $M_{n-H}\equiv m_{n}-m_{H}=782.27$~keV is the mass difference between the neutron and the hydrogen atom.
With Eqs.~\eqref{eq:Q} and \eqref{eq:Ex}, the end-point energy is given as 
\begin{equation}
Q_{\beta}-\mathcal{E}_{c,\gamma}^*=\lambda_{n}-\lambda_{p}+\Delta M_{n-H}-\mathcal{E}_{c,\lambda}.
\label{eq:epe}
\end{equation}
The correction energy $E_{\rm corr}$ is estimated from the fact that the ground state of the odd-mass nucleus corresponds to one quasi-particle state on the top of the even-mass nucleus~\cite{RingandSchuck}. 
The explicit form of $E_{\rm corr}$ is given as~\cite{Minato2021}:
\begin{eqnarray}
E_{\rm corr}&=
\begin{cases}
 E_{p_0}+E_{n_0}& (\text{$\beta$-decay for even-even nucleus})\\
 E_{p_0} & (\text{for even-odd})\\
 E_{n_0} & (\text{for odd-even})\\
 0 & (\text{for odd-odd}),
\end{cases}
\label{eq:correct}
\end{eqnarray}
where $E_{p_{0}}$ and $E_{n_{0}}$ are the lowest quasiparticle energies for proton and neutron, respectively.
\par
The shape factor $C(W)$, which depends on the type of $\beta$-decay, $c$, is calculated in the same way as Ref.~\cite{Tomislav2016}, but the relativistic correction terms of $\xi'v$ and $\xi'y$ are reduced to the non-relativistic limit~\cite{Behrens1971}.
The ratio of weak axial-vector/vector coupling constants reads $g_A=-1.2762(5)$~\cite{PDG2020}.
However, we use $g_A=-1$ instead, to consider the couples to more complicated states~\cite{Menendez2011}, such as higher-order configurations and hadronic degree of freedom.
The quenching factor in this work is thus about $0.784$, which is consistent to other studies on the GT transitions and $\beta$ decays~\cite{Martinez-Pinedo1993, Nakada1996, Jokinen1998}.
\par
The transition strength of the external field operator $\hat{\mathcal{O}}_{c}$~\cite{Tomislav2016} is represented by 
\begin{equation}
    B_{c,\gamma}=\left|\sum_{pn}\langle p||\hat{\mathcal{O}}_{c}||n \rangle 
    \left(u_p v_n X_{pn}^{(c,\gamma)}+\eta u_n v_p Y_{pn}^{(c,\gamma)}\right)\right|^2,
\end{equation}
where $u$ and $v$ are the coefficients in the canonical representation of $U$ and $V$ wave functions of Eq.~\eqref{eq:UV} and $\eta$ is $+1 (-1)$ when $\hat{\mathcal{O}}_{c}$ is even (odd) under time reversal.

%
%
%
%
%

\subsection{Bayesian neural network}
\label{sect:BNN}
To obtain a better description for the isoscalar pairing strengths $V$, we adopt an approach of the BNN, which has been successfully applied to the predictions of nuclear masses~\cite{ZM2018} and $\beta$-decay half-lives~\cite{ZM2019}. In the BNN approach, the model parameters $\boldsymbol{\omega}$ in the neural network are described by the posterior distribution $p(\boldsymbol{\omega}|D)$, 
\begin{equation}
 p(\boldsymbol{\omega}|D)=\frac{p(D|\boldsymbol{\omega})p(\boldsymbol{\omega})}{p(D)},
\end{equation}
where $p(D)$ is a normalization constant. The learning data are $D=\{(\boldsymbol{x}_1,V_1),(\boldsymbol{x}_2,V_2),...,(\boldsymbol{x}_N,V_N)\}$, where $V_k$ is the optimized isoscalar pairing strength of nucleus $\boldsymbol{x}_k=(Z, N)_k$. The prior distribution $p(\boldsymbol{\omega})$ is set as a Gaussian distribution with zero mean. The conditional probability is $p(D|\boldsymbol{\omega}) = \exp(\chi^2/2)$ with 
\begin{equation}
 \chi^2=\sum_{n=1}^N\left[\frac{S(\boldsymbol{x};\boldsymbol{\omega})-V_k}{\Delta V_k}\right]^2,
\end{equation}
where $\Delta V_k$ is the noise error, and the inverse of its square $1/\Delta V_k^2$ is set to a gamma distribution as in Ref.~\cite{ZM2018}. The function $S(\boldsymbol{x};\boldsymbol{\omega})$ is described by a neural network with one hidden layer, i.e.,
\begin{equation}
 S(\boldsymbol{x};\boldsymbol{\omega})=a+\sum_{j=1}^H b_j \tanh\left(c_j+\sum_{i=1}^I d_{ji}x_i\right).
\end{equation}
So the parameters of neural network are $\boldsymbol{\omega}=\{a,b_j,c_j,d_{ji}\}$. The number of hidden neurons is taken as $H=30$ in this work. With the specified $p(\boldsymbol{\omega})$ and $p(D|\boldsymbol{\omega})$, $p(\boldsymbol{\omega}|D)$ can be sampled using the Markov chain Monte Carlo algorithm. The prediction and the corresponding uncertainty of $S(\boldsymbol{x};\boldsymbol{\omega})$ for any input $(Z,N)$ are then calculated by its mathematical expectation and standard deviation on $p(\boldsymbol{\omega}|D)$.
\subsection{Technical notes for systematical calculation of half-lives}
\label{sect:tech}
For the systematical calculation of $\beta$-decay half-lives in the present framework, we sometime confront a problem of a phase transition.
This transition occurs when the correlated ground-state energy of $pn$QRPA is lower than the HFB ground-state energy.
Because the $pn$QRPA is the model that assumes a small-amplitude oscillation around the energy minimum for a collective coordinate, the $pn$QRPA equation of Eq.~\eqref{eq:QRPA} has an instability solution when this transition occurs (c.f. Sect.~8.4.2 of Ref.~\cite{RingandSchuck}). 
In the present framework, the phase transition is triggered when the isoscalar spin-triplet residual interaction is too strong and the first $1^+$ state is lower than the ground state calculated by the HFB.
For such a case, we switch to the proton-neutron quasiparticle Tamm-Dancoff approximation ($pn$QTDA), namely set the backward amplitudes, $B$ of Eq.~\eqref{eq:QRPA}, to be zero matrix, and omit the ground-state correlation incorporated by the $pn$QRPA.
In general, the result of $pn$QTDA is quite similar to that of $pn$QRPA for neutron-rich nuclei because the backward amplitudes of $pn$QRPA are appreciably hindered due to a large difference of the Fermi energies between proton and neutron.
We will discuss the influence of this problem later.

%
%
%
%
%
%
%

\section{Results and discussion}
\label{sect:result}
\subsection{Isovector pairing and model-space dependence}
We first study the relation of the zero-range isovector pairing force and the $\beta$-decay half-life by varying $\varepsilon_{\rm cut}$ that is introduced for the cutoff energy of HFB equation of Eq.~\eqref{eq:HFB}.
To remove the contribution from the isoscalar pairing strength, we discuss in this section by setting the strength $V=0$ of Eq.~\eqref{eq:isp}.
\par
\begin{figure}
    \centering
    \includegraphics[width=0.98\linewidth]{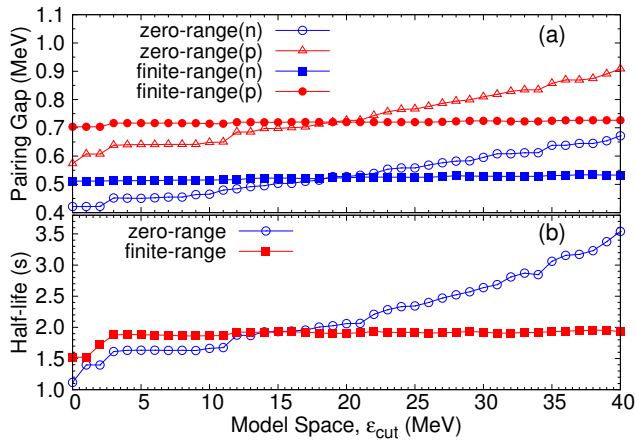}
    \caption{(a) Average pairing gaps of proton ($p$) and neutron ($n$) and (b) $\beta$-decay half-life of $^{128}$Cd as a function of the cutoff energy $\varepsilon_{\rm cut}$.}
    \label{fig:Model1}
\end{figure}

Figure~\ref{fig:Model1} shows the average proton and neutron pairing gaps weighted by the pairing density in the canonical basis, $\langle uv\Delta_q\rangle$~\cite{Sauvage1981, Bender2003} (top panel), and the $\beta$-decay half-life (bottom panel) of $^{128}$Cd as a function of $\varepsilon_{\rm cut}$.
The calculations are carried out with the D1S finite-range force of Eq.~\eqref{eq:finiteforce} or the zero-range volume type force given by
\begin{equation}
V_{\delta,q}^{(1)}(\vb*{r}_1,\vb*{r}_2)=-V_{\delta,q}\delta(\vb*{r}_1-\vb*{r}_2).
\end{equation}
The proton and neutron pairing strengths of the zero-range force are $V_{\delta,p}=194$ and $V_{\delta,n}=173$~MeV, respectively, which are determined so that the average pairing gaps are equal to those of the finite-range force when $\varepsilon_{\rm cut}=20$ MeV.
For the pairing gaps shown in Fig.~\ref{fig:Model1}(a), the zero-range force shows a little increment from $\varepsilon_{\rm cut}=0$ to $3$~MeV and a plateau from $\varepsilon_{\rm cut}=3$ to $10$~MeV.
Above $\varepsilon_{\rm cut}=10$~MeV, the gaps both for proton and neutron start to monotonically increase with the model space.
This model-space dependence is consistent to what is reported for the nuclear matter~\cite{Takahara1994}. 
On the other hand, the pairing gaps for the finite-range force are rather insensitive to the model space, although very small increments are observed.
\par
The model-space dependence seen in the pairing gaps affects the $\beta$-decay half-life.
It should be noted that the model-space dependence of half-life is more complicated than that of the pairing gaps, because the half-life also depends on the two-quasiparticle model space of the $pn$QRPA and the Fermi energies that are dependent on $\varepsilon_{\rm cut}$ as well.
The result is shown in Fig.~\ref{fig:Model1}(b).
By $\varepsilon_{\rm cut}=3$~MeV, the half-life increases for both the zero-range and finite-range forces due to the enlargement of two-quasiparticle model space and the variations of the pairing gaps.
In the range of $3 \le \varepsilon_{\rm cut} \le 10$~MeV, the half-life is insensitive to the model space because the pairing gaps are almost constant and the low-lying states relevant to the $\beta$ decay are not sensitive to the number of enlarging two-quasiparticle model space, which have higher energies.
The half-life of the zero-range pairing starts to increase from $\varepsilon_{\rm cut} \sim 12$~MeV.
This is mainly because the end-point energy in Eq.~\eqref{eq:epe} is inversely proportional to $\mathcal{E}_{c,\lambda}$, the eigenvalues of the $pn$QRPA, which grow up with increasing the two quasiparticle energies and the pairing gaps.
For example, the end-point energies are $3.30$~MeV for $\varepsilon_{\rm cut}=20$~MeV and $2.93$~MeV for $\varepsilon_{\rm cut}=40$~MeV in the case of zero-range force.
Note that this variation is at most the magnitude of $\Delta_{p}+\Delta_{n}$.
On the other hand, the half-life calculated by the finite-range force is rather insensitive to the model space, because the quasiparticle energy does not change significantly and so does the end-point energy.
The end-point energies are $3.35$ and $3.33$~MeV for $\varepsilon_{\rm cut}=20$ and $40$~MeV in the case of finite-range force, respectively.
The variation of the Fermi energies in the first and second terms in Eq.~\eqref{eq:epe} does not affect the half-life significantly.
They are canceled out by the Fermi energies included in the QRPA phonon energy $\mathcal{E}_{c,\lambda}$, which is expressed in the limit of the non-interacting particle model by $\mathcal{E}_{c,\lambda} \simeq |\lambda_{n}-\varepsilon_{n}|+|\varepsilon_{p}-\lambda_{p}|$.
\begin{figure}
    \centering
    \includegraphics[width=0.96\linewidth]{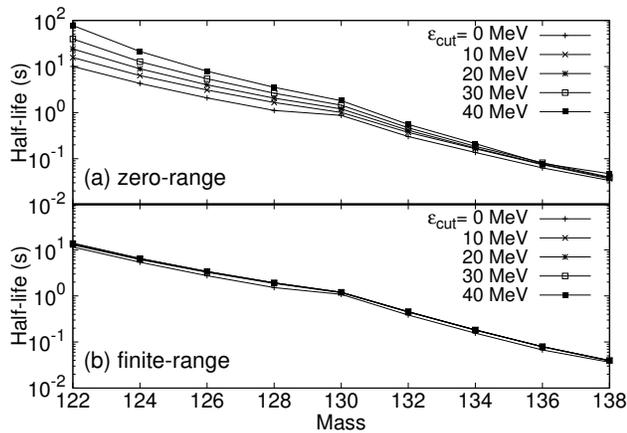}
    \caption{Half-lives of Cd isotopes calculated with different model space $\varepsilon_{\rm cut}$ for (a) the zero-range and (b) the finite-range isovector pairing interactions.}
    \label{fig:Model2}
\end{figure}
\par
We further study the isotope dependence of $\beta$-decay half-lives in the Cd ($Z=48$) isotopes with the zero-range and finite-range isovector pairing forces.
Figure~\ref{fig:Model2} shows the half-lives of Cd isotopes calculated with the model spaces, $\varepsilon_{\rm cut}=0, 10, 20, 30, 40$~MeV, where the top and bottom panels are the results of the zero-range and finite-range forces, respectively.
For the zero-range force, the half-lives of different $\varepsilon_{\rm cut}$ deviate largely.
The deviations of half-lives are sizable for the light-mass isotopes and become smaller with increasing mass number.
The variation of end-point energy, {\it i.e.}, the variation of pairing gaps, is only a few hundred keV, and its effect on half-life is significant when the end-point energy is small, while it diminishes for the neutron-rich nuclei that have a much larger end-point energy than the variation of pairing gaps.
In contrast, the half-lives calculated with the finite-range force converge rapidly in a small $\varepsilon_{\rm cut}$ for all the isotopes due to the insensitivity of the pairing gaps to the size of model space.
\par
In practice, a systematical calculation of $\beta$-decay half-lives is carried out by fixing $\varepsilon_{\rm cut}$ to a certain value. 
However, it is anticipated that the calculated half-lives will have different behaviors of isotopic dependence on the chosen size of model space in response of the pairing gap variation and $Q_{\beta}$.
This fact will not be favorable for a systematical estimation of the isoscalar pairing strength.
In contrast, the half-lives calculated with the finite-range pairing forces are rather insensitive to choice of the model space.
Such a result has an outstanding advantage of reducing the uncertainties arising from the selection of model space and is the main reason that the finite-range force for both the isoscalar and isovector pairing channels is adopted in this work.
In the following calculations, We adopt $\varepsilon_{\rm cut}=40$~MeV to ensure including enough configurations and to satisfy the Ikeda sum-rule by more than $99.8\%$ for the GT transition.
\subsection{Isoscalar pairing strength}
\label{sect:isoscalar}
The $\beta$-decay half-lives are also sensitive to the isoscalar spin-triplet pairing in addition to the isovector spin-singlet pairing.
The isoscalar pairing strength is independent of other effective forces in the present framework and is freely adjusted in calculating half-lives.
The isoscalar pairing strength that reproduce the experimental half-life is different for different nuclei, and an isospin-dependent force has been used for the previous systematical calculations of half-lives~\cite{Niu2013,Wang2016,Tomislav2016}.
However, it is still not clear whether the isospin-dependent force is an appropriate form.
In this section, we study the relation of the isospin pairing strength and the half-life more carefully.
\par
To quantify the prediction performance of an isospin pairing strength, let us begin with defining the mean deviation of the calculated and experimental half-lives for $N$ nuclei as
\begin{equation}
\bar{r}=\frac{1}{N} \sum_{i}^{N} r_{i},\quad r_{i}=\log_{10}\left(\frac{T_{{\rm calc},i}}{T_{{\rm exp},i}}\right),
\label{eq:r}
\end{equation}
and the standard deviation as
\begin{equation}
s=\sqrt{\frac{1}{N} \sum_{i}^{N} r_{i}^{2}}.
\end{equation}
Note that they are defined in the logarithmic scale considering the wide range of half-lives of unstable nuclei.
The ideal condition is $\bar{r}=s=0$.
We use the evaluated data of NUBASE2016~\cite{NUBASE16} for $T_{\rm exp}$ and choose the nuclei that have $T_{\rm exp} < T_{\rm exp}^{\rm max}$ with the upper limit of $T_{\rm exp}^{\rm max}$, if the computed $Q_{\beta}$ of Eq.~\eqref{eq:Q} is greater than $0$.
Moreover, if a half-life is insensitive to the isoscalar pairing strength, the nucleus is excluded from the chosen group.
\subsubsection{Single isoscalar pairing strength}
\label{sect:single}
\begin{figure}
    \centering
    \includegraphics[width=0.96\linewidth]{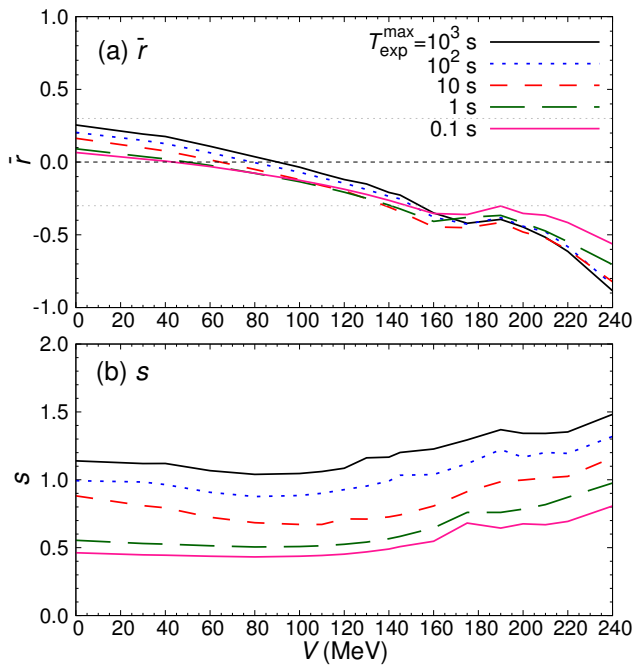}
    \caption{(a) Mean deviation $\bar{r}$ from the experimental data of $950$ nuclei and (b) standard deviation $s$ as a function of the isoscalar pairing strength $V$. The results of different ranges of $T$ are plotted. The dotted lines of $\bar{r}=\pm 0.3$ are also depicted.}
    \label{fig:TofV}
\end{figure}
We first investigate the mean deviation and the standard deviation by varying the isoscalar pairing strength.
Figure~\ref{fig:TofV} shows the results of $\bar{r}$ (top) and $s$ (bottom) for the isoscalar pairing strength in the range of $0\le V \le 240$~MeV, where different $T_{\rm exp}^{\rm max}$ are separately plotted.
For $V=0$~MeV, the calculated half-lives tend to be longer than the experimental data, giving $\bar{r}>0$ for all $T_{\rm exp}^{\rm max}$.
Increasing $V$, $\bar{r}$ gradually decrease and turn to be negative at some point.
Such turning points are different for different $T_{\rm exp}^{\rm max}$, and become smaller with shorter $T_{\rm exp}^{\rm max}$.
For $T_{\rm exp}^{\rm max} = 10, 1, 0.1$~s, in the range of $0 \le V \le 140$~MeV, $|\bar{r}|$ have small values less than $\sim 0.3$, which is equivalent to reproducing all the half-lives within about a factor of $2$ on average.
\par
The standard deviation $s$ shown in the bottom panel exhibits a rather weak dependence on $V$.
For $T_{\rm exp}^{\rm max} = 10^{3}$~s, $s \approx 1.1$, and $s$ becomes smaller as $T_{\rm exp}^{\rm max}$ is set to be shorter.
For $T_{\rm exp}^{\rm max} = 1, 0.1$~s, $s \approx 0.5$ with $0\le V \le 140$~MeV, which indicates that the present $pn$QRPA gives the half-lives within about a factor of $10^{0.5}\simeq 3.1$ on average with the single isoscalar pairing strengths.
\par
\begin{figure}
    \centering
    \includegraphics[width=0.96\linewidth]{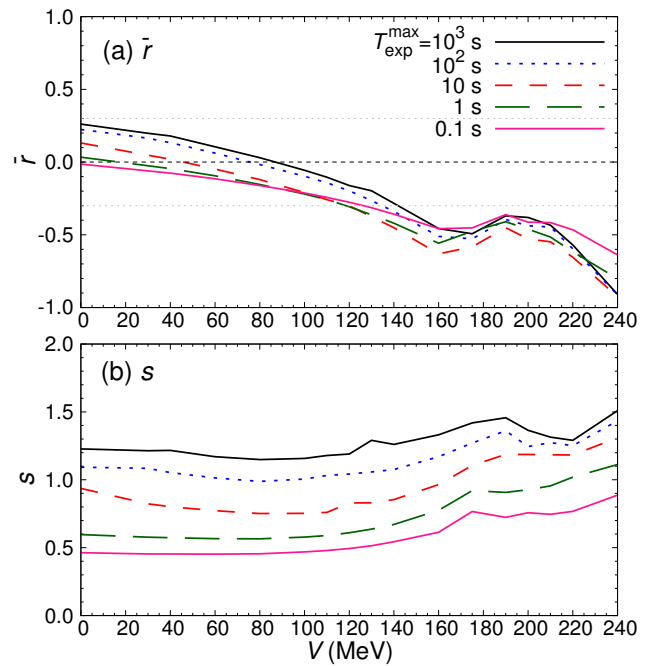}
    \caption{Same as Fig.~\ref{fig:TofV}, but for the $Z\ge 20$ nuclei with the quadrupole deformation parameter $\beta_{2} < 0.15$ and all nuclei with $Z<20$. The total number of target nuclei is $485$.}
    \label{fig:TofV2}
\end{figure}
The theoretical model of this work assumes the nuclear shape to be spherical, while a lot of nuclei have a deformed shape actually.
Therefore, we also study the effect of nuclear deformation. 
We again consider $\bar{r}$ and $s$ focusing only on the nuclei that have a small quadrupole deformation. 
We take the information on the quadrupole deformation parameter $\beta_{2}$ from Ref.~\cite{Ney2020}, in which the same SkO' functional~\cite{Reinhard1999} as this work is used and the data for the nuclei with $Z\ge20$ are available.
Figure~\ref{fig:TofV2} shows the results of $\bar{r}$ (top) and $s$ (bottom) of the nuclei with $Z\ge 20$, $\beta_{2}<0.15$ and all nuclei with $Z<20$, the total number of target nuclei being $485$.
By comparing Fig.~\ref{fig:TofV2} with Fig.~\ref{fig:TofV} that considers all the nuclei, we can learn the deformation effect on $\bar{r}$ and $s$ for nuclei with $Z\ge20$.
As compared with Fig.~\ref{fig:TofV}, the curves of $\bar{r}$ do not change significantly for all $T_{\rm exp}^{\rm max}$ although some variations are found for $T_{\rm exp}^{\rm max}=10^{3}, 10^{2}, 10$~s around $V=180\sim240$~MeV.
As a result, we obtain $|\bar{r}|<0.3$ within the almost same range of $V$ as Fig.~\ref{fig:TofV}.
Similarly, the curves of $s$ do not change significantly except variations around $V\simeq200$~MeV for $T_{\rm exp}^{\rm max}=10^{3}$~s.
As a consequence, nuclear deformation is important if the half-life is long and $V$ is large, while its effect becomes weakened if the half-life is short, at least for nuclei with $Z\ge20$.
This fact motivates us to predict the $\beta$-decay half-lives of neutron-rich nuclei, keeping the assumption of spherical shape of nuclei.
\subsubsection{Optimized isoscalar pairing strengths}

\begin{figure}
    \centering
    \includegraphics[width=1.0\linewidth]{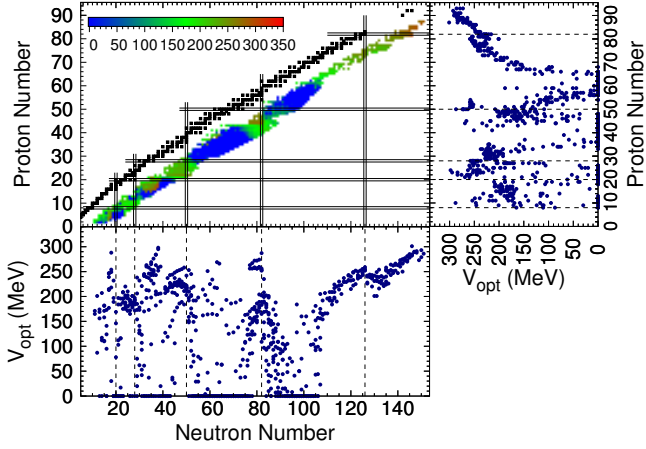}
    \caption{Optimized isoscalar pairing strengths $V_{\rm opt}$ in the $N$-$Z$ plane determined so as to reproduce the $\beta$-decay half-lives of NUBASE2016~\cite{NUBASE16}, and the corresponding projections to the $N$ and $Z$ axes. 
    The calculation is performed by the HFB+$pn$QRPA. Neutron and proton magic numbers are shown by the double and dashed lines.}
    \label{fig:VofN}
\end{figure}

We next seek the isoscalar pairing strength that reproduces experimental half-life for each nucleus.
We denote the strength as $V_{\rm opt}$.
The strength $V_{\rm opt}$ reflects two features: one is the isoscalar pairing strength itself, and another is a compensation of the missing nuclear structures that the present framework cannot describe.
Figure~\ref{fig:VofN} shows the result of $V_{\rm opt}$ in the $N$-$Z$ plane and the corresponding projections to the $N$ and $Z$ axes, where the magic numbers are drawn by the double and dashed lines.
Looking at the result of the $N$-$Z$ plane, on the one hand, we can observe that $V_{\rm opt}$ become high around the magic numbers and this structure is confirmed more clearly in the panels of the projections to the $N$ and $Z$ axes. 
On the other hand, the nuclei between the magic numbers, especially those around $(Z,N)=(40,70)$ and $(60,100)$, have small $V_{\rm opt}$.
As discussed in the next section, the nuclear deformation is particularly important around those regions~\cite{Stoitsov2003, AMEDEE, MassExplorer, InPACS,Ney2020} and we consider that its effect emerges through $V_{\rm opt}$.
In contrast, for the heavy nuclei above $Z=82, N=126$, $V_{\rm opt}$ stay high and no substantial decrease is found although most of those nuclei are deformed.
We consider that such a different behavior from that in the  $(Z,N)=(40,70)$ and $(60,100)$ regions is due to manifestation of the FF transitions, which are the main components of half-lives for the heavy nuclei above $(82, 126)$, and due to the weakening of the contribution from the GT transitions.
Because the FF transitions are less sensitive to the isoscalar pairing strength, the isoscalar pairing strengths stay high.
\par
The isospin-dependent isoscalar pairing force has been introduced in calculating the $\beta$-decay half-lives systematically in some previous works~\cite{Niu2013, Wang2016, Tomislav2016}.
However, we find that the isospin dependence in existing literature cannot represent $V_{\rm opt}$.
It is difficult to find a simple analytic function that expresses $V_{\rm opt}$ in spite of its characteristic structure found in Fig.~\ref{fig:VofN}.
For this reason, we apply the BNN for the estimation of $V$ of neutron-rich nuclei.
The BNN learning is carried out with the $950$ data points in the $N$-$Z$ plane of Fig.~\ref{fig:VofN}.
\par
\begin{figure}
    \centering
    \includegraphics[width=0.99\linewidth]{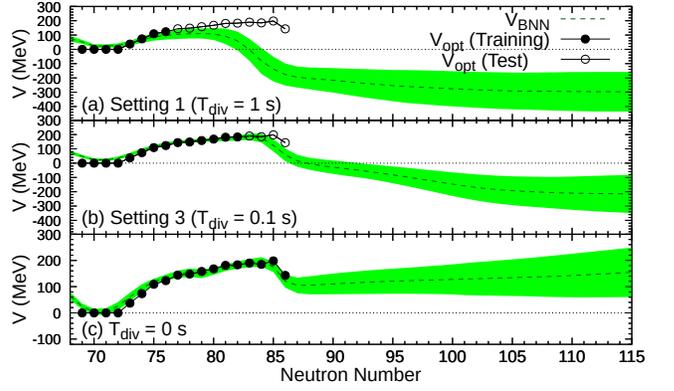}
    \caption{Isoscalar pairing strengths in the Cd isotopes estimated by the BNN ($V_{\rm BNN}$). The top, middle, and bottom panels are the results of setting 1 ($T_{\rm div}= 1$~s), 3 ($T_{\rm div}= 0.1$~s) of Table~\ref{tab:ttset}, and that using all data for training ({\it i.e.} $T_{\rm div}= 0$ s), respectively. The mean values of $V_{\rm BNN}$ are shown by the dashed lines and the shaded areas are the uncertainties of $1\sigma$. The data set of training $V_{\rm opt}$ (filled circle) and test $V_{\rm opt}$ (open circle) are plotted together. Evolution of $V_{\rm BNN}$ with increasing the number of training data can be seen through the panels (a)--(c).}
    \label{fig:BNN}
\end{figure}

As a typical example, the results of $V$ (denoted as $V_{\rm BNN}$) for the Cd isotopes are shown in Fig.~\ref{fig:BNN}(c), where the mean values are drawn by the dashed line and the $1\sigma$ uncertainty by the shaded area.
The BNN reasonably reproduces $V_{\rm opt}$, which are shown by the filled circles, and predicts $100 \le V \le 150$~MeV from $N=86$ to $115$.
Although the uncertainties become larger as going to the neutron-rich side, we have checked that its propagation to the half-life uncertainties is not significant.
As an example, $T_{\rm calc}=0.789\pm0.007$~ms for $^{163}$Cd.
\par
We consider that $V_{\rm BNN}$ need to be checked whether they are fair enough for the prediction of $\beta$-decay half-lives of the
unmeasured neutron-rich nuclei.
One of the methods to do so is to separate $T_{\rm exp}$ data used to obtain $V_{\rm opt}$ into the training and test sets.
The training set is used for calculating $V_{\rm BNN}$ as we have just done, and the test set is used for quantifying the predictive performance of the half-lives computed with the $V_{\rm BNN}$.
Here we divide the total data into the training set with $T_{\rm exp} \ge T_{\rm div}$ and the test set with $T_{\rm exp} < T_{\rm div}$, and calculate $\bar{r}$ and $s$ by taking $T_{\rm div} = 1.00$, $0.50$, $0.10$, and $0.05$~s as examples.
Table~\ref{tab:ttset} lists the result of $\bar{r}$ and $s$ for four different settings of the training and test data.
When the training data is limited to $T_{\rm exp} \ge 1$~s, we obtain $\bar{r}=-0.080$ and $s =0.478$.
For $T_{\rm div}=0.50$~s, $\bar{r}$ slightly improves becoming $-0.020$ and $s$ worsens slightly.
Increasing the number of training data further, $s$ becomes even smaller and the mean deviation fluctuates around $\bar{r}=-0.050$.
For $T_{\rm div}=0.05$~s, we obtain $\bar{r}=-0.031$ and $s=0.270$.
Figure~\ref{fig:BNN} illustrates the evolution of $V_{\rm BNN}$ for the Cd isotopes with increasing number of training data, where $V_{\rm BNN}$ of $T_{\rm div}=1$, $0.1$~s and that using all data for training are shown.
For $T_{\rm div}=1$~s, the BNN severely underestimates the test data of $V_{\rm opt}$.
We consider that the number of training data is not enough for the prediction of $V$ in this case.
However, the predictive performance of BNN gradually improves with increasing number of training data.
From Fig.~\ref{fig:BNN}(b), the BNN reproduces the test data of $V_{\rm opt}$ roughly within the uncertainty when $T_{\rm div}=0.1$~s, and accordingly the standard deviation $s$ improves from that of $T_{\rm div}=1.0, 0.5$~s, as found in Table~\ref{tab:ttset}. 
As mentioned above, the BNN reasonably reproduces $V_{\rm opt}$ when we use all experimental data (Fig.~\ref{fig:BNN}(c)).
\par
We find that $V_{\rm BNN}$ become negative at some point for Fig.~\ref{fig:BNN}(a) and (b) despite that $V$ is a positive number by definition.
This issue is also observed for elements other than Cd.
This indicates that the BNN does not work well for very neutron-rich sides if the number of training data set is small.
However, we found that the number of negative $V_{\rm BNN}$ greatly decreases with increasing number of training data.
As seen from Fig.~\ref{fig:BNN}(c), the issue of negative $V_{\rm BNN}$ is reasonably solved, showing a $V\simeq150\pm90$~MeV at $N=115$.
We thus consider that the number of training data that are taken from presently available experimental data is adequate for the prediction of $V$ of neutron-rich nuclei.
\par
From Table~\ref{tab:ttset}, when $T_{\rm div} = 0.05$~s (the number of training data is $841$), we obtained $\bar{r}=-0.031$ and $s=0.270$.
Transforming them in the linear scale, $\bar{r}_{\rm lin} \simeq 0.93$ and $s_{\rm lin} \simeq 1.9$.
The same performance is expected in the predicted half-lives when we fully use experimentally available data as the training set.
In the latter section, the predicted half-lives of $V_{\rm BNN}$ will be further validated by comparing with new experimental data.
\begin{table}
\caption{Training and test data sets, which are selected for $T_{\rm exp} \ge T_{\rm div}$ and $T_{\rm exp} < T_{\rm div}$ from $950$ nuclei, respectively. Results of the mean deviation $\bar{r}$ and standard deviation $s$ using four different settings of training and test data are shown.}
\centering
\label{tab:ttset}
\begin{tabular}{c|c|cc|dd}
\hline\hline
 & $T_{\rm div}$ & Number of & Number of \\
Setting & (s) & training data & test data & \bar{r} & s \\
\hline
1 & $1.00$ & 569 & 381 & -0.080 & 0.478 \\
2 & $0.50$ & 626 & 324 & -0.020 & 0.494 \\
3 & $0.10$ & 776 & 174 & -0.085 & 0.335 \\
4 & $0.05$ & 841 & 109 & -0.031 & 0.270 \\
\hline
\end{tabular}
\end{table}
\subsection{Systematical calculation of $T$ with $V_{\rm BNN}$}
\label{sect:sys}
In this section, we present the result of the systematical calculation of $T$ with $V_{\rm BNN}$ and compare it with other theoretical data.
The targets are unstable nuclei against $\beta^{-}$ decay with the theoretically calculated two-neutron separation energy $S_{2n}>0$~MeV.
\par
\begin{figure}
    \centering
    \includegraphics[width=1.0\linewidth]{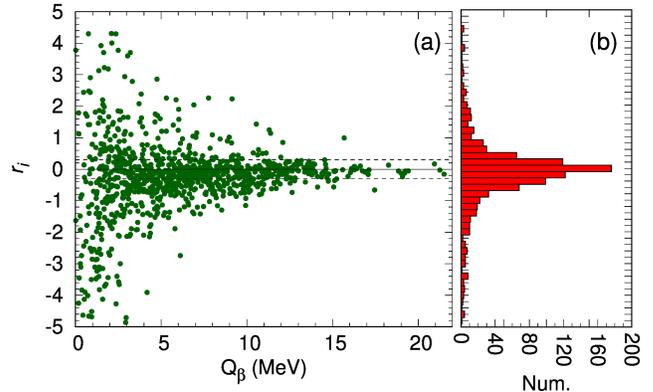}
    \caption{(a) Distribution of $r_{i}$ as a function of $Q_{\beta}$. (b) The histogram of the statistics of $r_{i}$ with a bin $\Delta r_{i}=0.2$. The mean deviation and the standard deviation of the histgram are $\bar{r}=-0.155$ and $s=1.153$, respectively.}
    \label{fig:R2Q}
\end{figure}

Figure~\ref{fig:R2Q}(a) shows $r_{i}$ as a function of $Q_{\beta}$ setting $T_{\rm exp}^{\rm max} = 10^{10}$~s.
The dashed lines indicate $r_{i}=\pm0.3$ that corresponds to reproduce $T_{\rm exp}$ within a factor of $2$.
At small $Q_{\beta}$ close to zero, $r_{i}$ distribute widely from $-5$ to $5$.
The $\beta$-decay half-life is sensitive to $Q_{\beta}$ and approximately proportional to $Q_{\beta}^{-5}$.
The EDFs including the present framework cannot always reproduce $Q_{\beta}$ with an accuracy of keV order, the wide fluctuation thus emerges at small $Q_{\beta}$.
With increasing $Q_{\beta}$, $r_{i}$ converge to around $0$.
This may imply that the present framework has a higher performance as going to the neutron-rich nuclei that have large $Q_{\beta}$.
The histogram shown in Fig.~\ref{fig:R2Q}(b) is the statistics of $r_{i}$ with a bin size $\Delta r_{i}=0.2$.
We can see that the calculated ratios distribute centering $r_{i}=0$.
The mean deviation and the standard deviation of the histogram are $\bar{r}=-0.155$ and $s=1.153$, respectively.
\par
\begin{figure}
    \centering
    \includegraphics[width=1.0\linewidth]{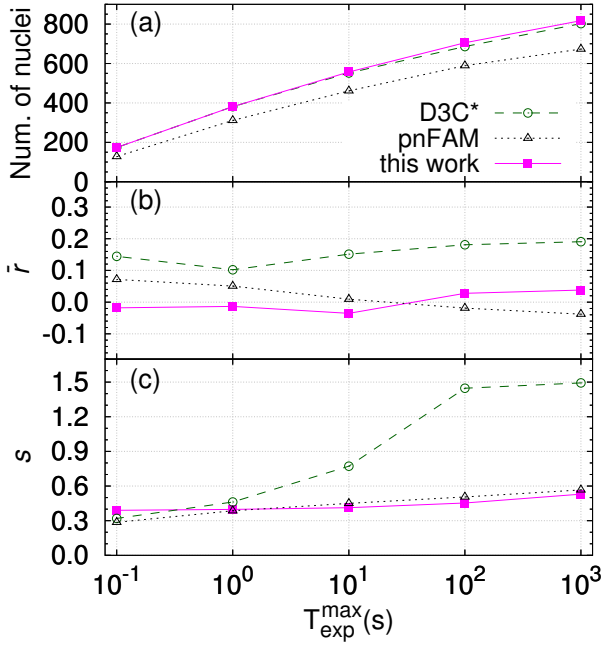}
    \caption{(a) Number of nuclei, (b) mean deviation $\bar{r}$, and (c) standard deviation $s$ within the upper limit of half-life $T_{\rm exp}^{\rm max}$. The results of Skyrme HFB + $pn$QRPA with $V_{\rm BNN}$ (this work), D3C$^*$~\cite{Tomislav2016}, and pnFAM~\cite{Ney2020} are plotted together for comparison.}
    \label{fig:rs}
\end{figure}

Since it is difficult to reproduce a long $T_{\rm exp}$ in the present framework and our interest is weighted on the prediction of shorter half-lives rather than that of longer ones, we present the results limiting $T_{\rm exp}^{\rm max}$ to a relatively short time. 
The mean deviation $\bar{r}$ and standard deviation $s$ for $T_{\rm exp}^{\rm max}=10^{-1}, 10^{0}, 10^{1}, 10^{2}$, and $10^{3}$~s are shown in Fig.~\ref{fig:rs}, where the results of D3C$^*$~\cite{Tomislav2016} and $pn$FAM~\cite{Ney2020} are also plotted together for comparison. 
The top panel shows the numbers of nuclei within the range of the given upper limits of $T_{\rm exp}^{\rm max}$.
This work includes almost the same numbers of nuclei as the D3C$^*$ for the present analysis, while those of $pn$FAM are smaller because the $Z<20$ nuclei are not considered there.
The middle panel shows the results of $\bar{r}$.
This work provides $|\bar{r}|$ within $0.1$ for different $T_{\rm exp}^{\rm max}$ and is comparable with the $pn$FAM, while the D3C$^*$ gives much larger values not only for long $T_{\rm exp}^{\rm max}$ but also for short $T_{\rm exp}^{\rm max}$.
The bottom panel illustrates the results of $s$.
This work shows that $s$ are within $0.6$ even for $T_{\rm exp}^{\rm max}=10^{3}$~s and gradually decrease with shorter $T_{\rm exp}^{\rm max}$. 
This work is almost comparable with the $pn$FAM for $T_{\rm exp}^{\rm max} = 10^{0}, 10^{1}$~s and gives a slightly larger value than the $pn$FAM and D3C$^*$ by about $0.1$ for $T_{\rm exp}^{\rm max}=10^{-1}$ s.
%
\par
Table~\ref{tab:all} shows the mean variation and standard deviation when nuclei are categorized to the even-even, even-odd, odd-even, and odd-odd ones for $T^{\rm max}_{\rm exp}=10$~s.
For the even-even and even-odd nuclei, $\bar{r}$ of this work is better than the $pn$FAM and comparable with the D3C$^{*}$.
For the odd-even nuclei, this work is better than the D3C$^{*}$. For the odd-odd nuclei, this work is better than both the D3C$^{*}$ and the $pn$FAM.
For the standard deviation $s$, this work is better than the D3C$^*$ for all the categories.
This work is also better than the $pn$FAM for the even-even and even-odd nuclei and comparable for the odd-odd nuclei.
Only for the odd-even nuclei, the $pn$FAM clearly shows a better performance of $\bar{r}$ and $s$ than this work.
\begin{table}
    \centering
    \caption{Mean deviation $\bar{r}$ and standard deviation $s$ grouped by the even-even (E-E), even-odd (E-O), odd-even (O-E), and odd-odd (O-O) nuclei for $T^{\rm max}_{\rm exp}=10$~s. The results of the $pn$QRPA calculations with $V_{\rm BNN}$ (this work), D3C$^*$~\cite{Tomislav2016}, and $pn$FAM~\cite{Ney2020} are compared.}
    \begin{tabular}{r|dd|dd|dd}
    \hline\hline
            & \multicolumn{2}{c|}{This work} &
                \multicolumn{2}{c|}{D3C$^*$} & \multicolumn{2}{c}{$pn$FAM} \\
            & \bar{r} & s & \bar{r} & s & \bar{r} & s \\
    \hline
         E-E & -0.009 & 0.294 & -0.001 & 0.475 & -0.039 & 0.428 \\
         E-O & -0.020 & 0.301 &  0.019 & 0.544 & -0.055 & 0.428 \\
         O-E &  0.043 & 0.406 &  0.153 & 0.608 & -0.014 & 0.338 \\
         O-O &  0.106 & 0.552 &  0.378 & 1.154 &  0.120 & 0.557 \\
    \hline
    \end{tabular}
    \label{tab:all}
\end{table}
\par
\begin{figure}
    \centering
    \includegraphics[width=1.0\linewidth]{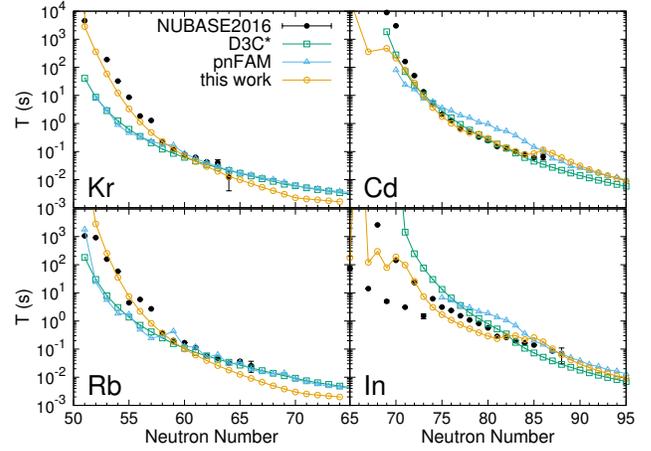}
    \caption{Calculated $\beta$-decay half-lives of this work, D3C$^*$~\cite{Tomislav2016}, $pn$FAM~\cite{Ney2020} for Kr, Rb, Cd, and In isotopes. The solid circles are taken from NUBASE2016~\cite{NUBASE16}.}
    \label{fig:half}
\end{figure}

Figure~\ref{fig:half} shows the $\beta$-decay half-lives of the Kr $(Z=36)$, Rb $(Z=37)$, Cd $(Z=48)$, and In $(Z=49)$ isotopes compared with the D3C$^*$ functional~\cite{Tomislav2016} and $pn$FAM~\cite{Ney2020}.
This work reproduces the experimental half-lives of nuclei from close-to-$\beta$-stability line to neutron-rich side reasonably.
This result is due to the flexible character of the BNN that does not assume a specific function for $V_{\rm opt}$. 
For the neutron-rich side where no experimental data is available, the present result is about half shorter than the $pn$FAM and $pn$RQRPA for the Kr and Rb isotopes, while close to the $pn$FAM for the Cd and In isotopes.
\par
\begin{figure}
    \centering
    \includegraphics[width=1.0\linewidth]{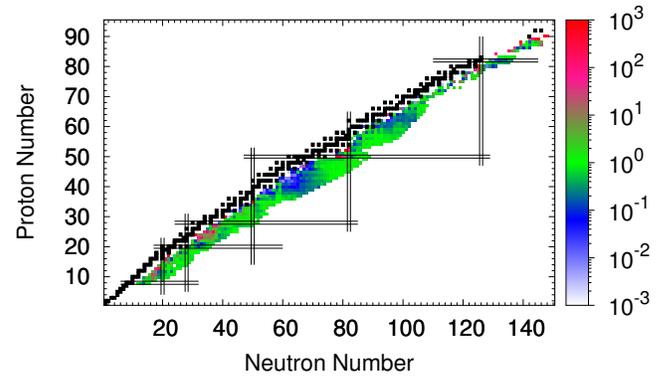}
    \caption{Ratios between the calculated and experimental half-lives in the $N$-$Z$ plane.}
    \label{fig:CoE}
\end{figure}
\begin{figure}
    \centering
    \includegraphics[width=1.0\linewidth]{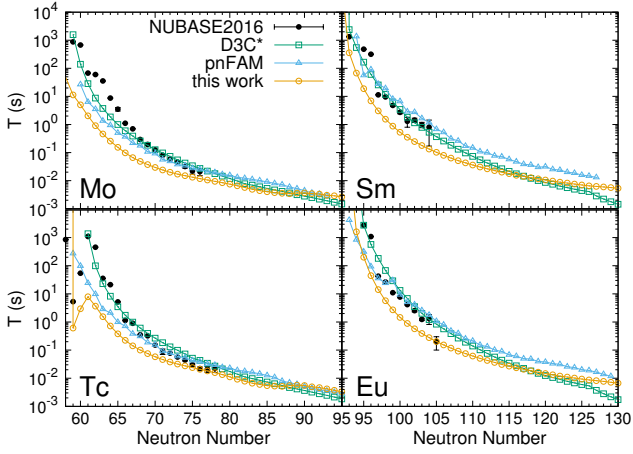}
    \caption{Same as Fig.~\ref{fig:half}, but for Mo, Tc, Sm, and Eu isotopes.}
    \label{fig:half_def}
\end{figure}

We also plot the ratios between the calculated and experimental half-lives in the $N$-$Z$ plane in Fig.~\ref{fig:CoE}.
We can see that the ratio is approximately $1$ for most nuclei.
However, underestimations are found around $(Z,N)=(45, 65)$ and $(65, 95)$.
As mentioned above, in these regions, the nuclear deformation plays a significant role~\cite{Stoitsov2003, AMEDEE, MassExplorer, InPACS}.
The low-lying states related to the $\beta$ decay are degenerate if one assumes the spherical shape.
The nuclear deformation breaks the degeneration and fragments the low-lying states into a wider energy range, resulting in a longer half-life than that with the spherical shape.
For many nuclei, the isoscalar pairing strength could effectively substitute the effect of nuclear deformation.
Figure~\ref{fig:half_def} shows the $\beta$-decay half-lives of the Mo $(Z=42)$, Tc $(Z=43)$, Sm $(Z=62)$, and Eu $(Z=63)$ isotopes, in which the nuclear deformation becomes significant.
This work clearly gives shorter half-lives than the experimental data for the light-mass nuclei.
The quadrupole deformation parameters are about $\beta_{2} \simeq 0.20$ for $N=60 \sim 70$ of the Mo and Tc isotopes and $\beta_{2} \simeq 0.34$ for $N=95 \sim 110$ of the Sm and Eu isotopes~\cite{Ney2020}.
Due to the large deformations, the isoscalar pairing strength calculated by the BNN could not substitute the effect.
On the other hand, the $pn$FAM that considers the nuclear deformation reproduces the half-lives of those nuclei reasonably.
It is reported that the half-life of $^{106}$Zr is increased by about a factor of $3$ if the nuclear deformation is considered~\cite{Yoshida2015}.
Therefore, the underestimations found in those nuclei are expected to be improved by considering the nuclear deformation.
%
\par
\begin{figure}
	\centering
	\includegraphics[width=1.0\linewidth]{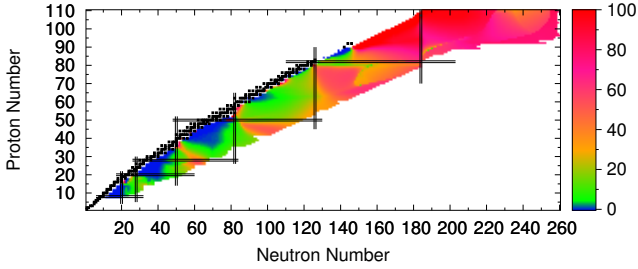}
	\caption{Percentages of the contributions from the first-forbidden transitions to the total $\beta$-decay rates.}
	\label{fig:FF}
\end{figure}

Figure~\ref{fig:FF} shows the percentages of the contributions from the first-forbidden transitions to the total $\beta$-decay rates.
As mentioned already, the FF transition becomes a main contributor of $\beta$ decay for the nuclei above $Z=82, N=126$.
In particular, its percentage becomes even higher when getting across $N=126$.
For $N<126$, the allowed transition is the main contributor of $\beta$ decay, however, we can see that the FF transition becomes important for some spots close to the $\beta$-stability line and around $Z=28, N=60$ and $Z=45, N=100$ regions, where the transitions from the neutron $sdg$ shell to the proton $pf$ shell and from the neutron $3p$, $2f$, and $1h$ orbitals to the proton $sdg$ shell are open, respectively.
\subsection{Comparison with new experimental data}
So far, our analysis of the isoscalar pairing strength is carried out based on the NUBASE2016~\cite{NUBASE16}.
It would be a concern whether the present work could predict $T_{\rm exp}$ if new experimental results that are not in the NUBASE2016 come out.
Recently, RIKEN measured the $\beta$-decay half-lives of $55$ neutron-rich nuclei of $Z=50\sim55$~\cite{RIKEN2020}.
Fourteen nuclei out of them are not in the NUBASE2016, which are $^{140\sim142}$Sb, $^{139\sim144}$Te,$^{143\sim146}$I, and $^{148}$Xe.
We use them for estimating the predictive performance of the present framework.
\par
\begin{figure}
    \centering
    \includegraphics[width=1.0\linewidth]{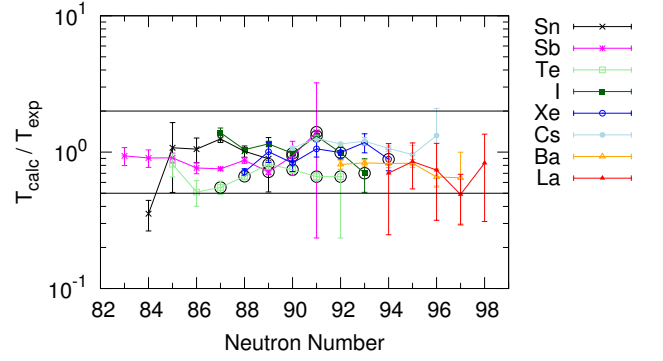}
    \caption{Ratios between the calculated $T_{\rm calc}$ by the Skyrme HFB+$pn$QRPA with BNN and the newly measured $T_{\rm exp}$ at RIKEN~\cite{RIKEN2020}. Data points of 14 nuclei that are not listed in the NUBASE2016~\cite{NUBASE16} are highlighted by the solid circles.}
    \label{fig:RIBF}
\end{figure}

The ratios between the half-lives calculated by the Skyrme HFB+$pn$QRPA and the new $55$ data of $T_{\rm exp}$ are shown in Fig.~\ref{fig:RIBF}.
Except $^{134}$Sn, this work can predict the new experimental data within a factor of $2$.
The half-lives of fourteen nuclei that are not in NUBASE2016 are also reproduced well, validating that the present approach is effective for prediction.
\par
\begin{figure}
    \centering
    \includegraphics[width=1.0\linewidth]{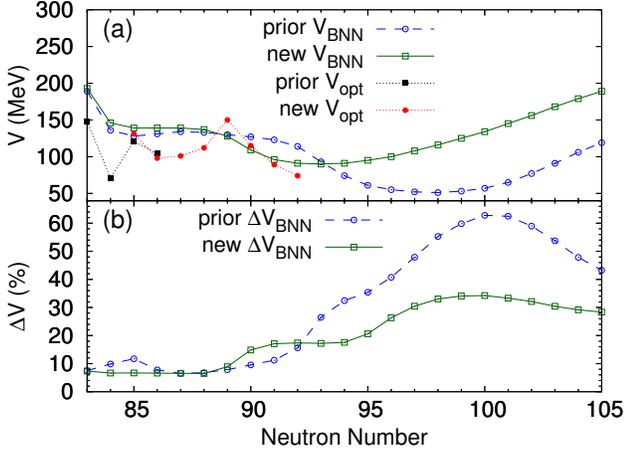}
    \caption{(a) Isoscalar pairing strengths $V$ and (b) their uncertainties $\Delta V$ for Te isotopes estimated by the BNN with the NUBASE2016~\cite{NUBASE16} (prior $V_{\rm BNN}$) and with the NUBASE2016 and new measurements~\cite{RIKEN2020} (new $V_{\rm BNN}$). The optimized strengths $V_{\rm opt}$ estimated with the NUBASE2016 and new data are also shown.}
    \label{fig:after1}
\end{figure}

Adding the newly measured half-lives to the training data of the BNN, we estimate new isoscalar pairing strengths and study the variations from the prior ones.
The number of training data is $964=950+14$ in total, in which the overlapping data are replaced by the new ones.
The results for the Te ($Z=52$) isotopes are shown in Fig.~\ref{fig:after1}, where the top and bottom panels are the isoscalar pairing strengths $V$ and the corresponding uncertainties $\Delta V$, respectively.
The ``prior'' and ``new'' $V_{\rm BNN}$ mean that calculated only with the NUBASE2016 and with both the NUBASE2016 and new experimental data, respectively.
The NUBASE2016 compiles the Te isotopes up to $N=86$, and the new measurement added the data up to $N=92$.
We can see that the prior $V_{\rm BNN}$ predicts the new $V_{\rm opt}$ fairly well, and is close to the result of the new $V_{\rm BNN}$ up to around $N=90$.
Beyond $N=90$, the prior $V_{\rm BNN}$ exhibits a different $N$-dependence from the new $V_{\rm BNN}$.
For $N=100$, the difference between the prior and new $V_{\rm BNN}$ is about $100$~MeV.
In Fig.~\ref{fig:after1}(b), the prior and new $\Delta V_{\rm BNN}$ show similar uncertainties up to $N=92$, and begin to show difference above $N=92$.
Due to the increment of data points, the uncertainties for the new $\Delta V_{\rm BNN}$ are significantly reduced, being smaller than those for the prior $\Delta V_{\rm BNN}$.
\par
\begin{figure}
    \centering
    \includegraphics[width=1.0\linewidth]{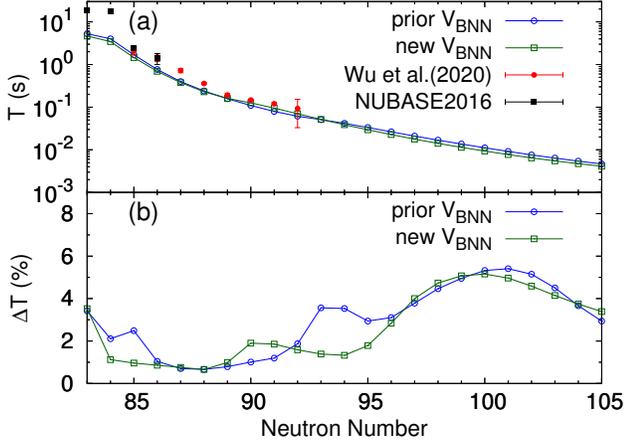}
    \caption{(a) $\beta$-decay half-lives and (b) their uncertainties of Te isotopes calculated with the isoscalar pairing strengths estimated by the BNN with the prior $V_{\rm BNN}$ and new $V_{\rm BNN}$. The experimental data of NUBASE2016~\cite{NUBASE16} and new measurements~\cite{RIKEN2020} are also plotted for comparison.}
    \label{fig:after2}
\end{figure}

The $\beta$-decay half-lives and the corresponding uncertainties calculated with the prior and new isoscalar pairing strengths are shown in Fig.~\ref{fig:after2}.
The uncertainty is calculate by
\begin{equation}
\begin{split}
\Delta T
&= \left| \frac{\partial T}{\partial V}\right| \Delta V
\sim \left| \frac{T(V)-T(V-\Delta V)}{V-(V-\Delta V)} \right| \Delta V\\
&=\left| T(V)-T(V-\Delta V)\right|.
\end{split}
\end{equation}
The calculated half-lives of the prior and new $V_{\rm BNN}$ show similar behaviors of isotopic dependence, although $V_{\rm BNN}$ of the neutron-rich side have a large difference as seen in Fig.~\ref{fig:after1}(a).
For example, the half-lives of $^{152}$Te ($N=100$) obtained with the prior and new $V_{\rm BNN}$ are $11.2$ and $9.3$~ms, respectively, showing about $20$\% difference.
In Fig.~\ref{fig:after2}(b), $\Delta T$ of the new $V_{\rm BNN}$ are meaningfully reduced for $N=93$, $94$, and $95$.
On the other hand, those for $N\ge96$ are almost the same as the prior $V_{\rm BNN}$.
The uncertainty is calculated with the multiplication of $\partial T/\partial V$ and $\Delta V$.
We confirmed that the new $\Delta V_{\rm BNN}$ around $N=100$ are about a half of the prior $\Delta V_{\rm BNN}$ as seen in Fig.~\ref{fig:after1}, while $\partial T/\partial V$ of the new $V_{\rm BNN}$ is about twice larger than that of the prior $V_{\rm BNN}$.
As a result, their $\Delta T$ become close to each other.
\par
Before closing this section, we should note the phase transition explained in Sect.~\ref{sect:tech}.
We observed $993$ nuclei that occur the phase transition when using only NUBASE2016 database as the training data.
For $Z\le68$, the corresponding nuclei are about $70$ and all close to the $\beta$-stability line.
No phase transition is observed for predicted nuclei for $Z\le68$, and therefore the influence of such a problem is very limited at least for those elements.
The remaining nuclei, about $93$\% of the phase transition, are in neutron-rich sides of $Z>68$, and the majority of them is nuclei with $Z\ge84$.
When adding new experimental data measured at RIKEN~\cite{RIKEN2020} to the training data, the number of nuclei showing the phase transition reduces to $552$, and no phase transition is observed for predicted nuclei with $Z\le72$.
From this result, we consider that the number of the phase transition tends to decrease with increasing the training data. 
We also consider that this problem can be remedied to some extent by considering the nuclear deformation because those correlations reduce the ground-state energy further.
In fact, many nuclei showing the phase transition are highly deformed.

\section{Summary and perspectives}
\label{sect:summary}
We studied the isoscalar pairing strength determined from the experimental $\beta$-decay half-lives of neutron-rich nuclei.
We first presented the model space dependence of half-lives for different types of isovector pairing force and showed the importance of using the finite-range force for the systematical investigation of $\beta$-decay half-lives.
\par
We next studied the mean deviation and standard deviation of half-life with different values of a single isoscalar pairing strength.
It was shown that the half-lives shorter than $10$~s were reproduced well within a limited range of the isoscalar pairing strength.
Limiting nuclei with a small deformation, the half-lives are reproduced within a factor of $2$ in the range of $0\le V \le 140$~MeV.
The isoscalar pairing strengths determined from the half-lives in NUBASE2016 showed a characteristic structure.
To represent the $N$ and $Z$ dependence, we applied the BNN and used it for the systematical prediction of half-lives.
We demonstrated that the calculated half-lives could predict the experimental data well by dividing the total data into the training and test sets.
\par
The calculated $\beta$-decay half-lives were compared with other models, showing a comparable result with $pn$FAM and a better one than D3C$^*$.
However, we found that some nuclei with a large deformation could not be well reproduced.
The calculated half-lives were also compared with the experimental data newly measured at RIBF in RIKEN.
It is found that the new data can be reproduced within a factor of $2$.
Using the new experimental data for the training set of BNN, we studied the variation of $V_{\rm BNN}$ and $T$.
Due to the new data points, $\Delta V_{\rm BNN}$ were reduced significantly and the uncertainties of $T$ for some nuclei also become smaller substantially.
We should note that the uncertainty discussed here accounts for the contribution only from the isoscalar pairing strength.
The present study considers the influences of new experiment on the prediction of half-lives.
However, to obtain a correct uncertainty of half-life within the present framework, it is required to also propagate the uncertainties of the parameters of SkO'~\cite{Reinhard1999} that originate from the bulk properties of nuclei.
\par
We expect that considering nuclear deformation will improve the predictive performance of the isoscalar pairing strengths and $\beta$-decay half-lives of neutron-rich nuclei, and plan to expand our framework to the axially deformed shape.
We also plan to calculate the $\beta$-delayed neutron emission and fission, which are also important for $r$-process simulation and nuclear data.
Although this work limited to the $\beta^{-}$ decays of neutron-rich nuclei, it is interesting to apply the formalism to the $\beta^{+}$ decays of neutron-deficient nuclei, too.
%
 \begin{acknowledgments}
 This research was conducted with the supercomputer HPE SGI8600 in the Japan Atomic Energy Agency. This work was supported by JSPS KAKENHI under Grant Nos.~21H01087, 18K13549, and 20H05648, the RIKEN Pioneering Project: Evolution of Matter in the Universe, and the National Natural Science Foundation of China under Grant No.~11875070.
 \end{acknowledgments}
\nocite{*}
\bibliography{biblio}

\begin{thebibliography}{51}%
\makeatletter
\providecommand \@ifxundefined [1]{%
 \@ifx{#1\undefined}
}%
\providecommand \@ifnum [1]{%
 \ifnum #1\expandafter \@firstoftwo
 \else \expandafter \@secondoftwo
 \fi
}%
\providecommand \@ifx [1]{%
 \ifx #1\expandafter \@firstoftwo
 \else \expandafter \@secondoftwo
 \fi
}%
\providecommand \natexlab [1]{#1}%
\providecommand \enquote  [1]{``#1''}%
\providecommand \bibnamefont  [1]{#1}%
\providecommand \bibfnamefont [1]{#1}%
\providecommand \citenamefont [1]{#1}%
\providecommand \href@noop [0]{\@secondoftwo}%
\providecommand \href [0]{\begingroup \@sanitize@url \@href}%
\providecommand \@href[1]{\@@startlink{#1}\@@href}%
\providecommand \@@href[1]{\endgroup#1\@@endlink}%
\providecommand \@sanitize@url [0]{\catcode `\\12\catcode `\$12\catcode
  `\&12\catcode `\#12\catcode `\^12\catcode `\_12\catcode `\%12\relax}%
\providecommand \@@startlink[1]{}%
\providecommand \@@endlink[0]{}%
\providecommand \url  [0]{\begingroup\@sanitize@url \@url }%
\providecommand \@url [1]{\endgroup\@href {#1}{\urlprefix }}%
\providecommand \urlprefix  [0]{URL }%
\providecommand \Eprint [0]{\href }%
\providecommand \doibase [0]{http://dx.doi.org/}%
\providecommand \selectlanguage [0]{\@gobble}%
\providecommand \bibinfo  [0]{\@secondoftwo}%
\providecommand \bibfield  [0]{\@secondoftwo}%
\providecommand \translation [1]{[#1]}%
\providecommand \BibitemOpen [0]{}%
\providecommand \bibitemStop [0]{}%
\providecommand \bibitemNoStop [0]{.\EOS\space}%
\providecommand \EOS [0]{\spacefactor3000\relax}%
\providecommand \BibitemShut  [1]{\csname bibitem#1\endcsname}%
\let\auto@bib@innerbib\@empty
\bibitem [{\citenamefont {Otsuka}\ \emph {et~al.}(2020)\citenamefont {Otsuka},
  \citenamefont {Gade}, \citenamefont {Sorlin}, \citenamefont {Suzuki},\ and\
  \citenamefont {Utsuno}}]{Otsuka2020}%
  \BibitemOpen
  \bibfield  {author} {\bibinfo {author} {\bibfnamefont {T.}~\bibnamefont
  {Otsuka}}, \bibinfo {author} {\bibfnamefont {A.}~\bibnamefont {Gade}},
  \bibinfo {author} {\bibfnamefont {O.}~\bibnamefont {Sorlin}}, \bibinfo
  {author} {\bibfnamefont {T.}~\bibnamefont {Suzuki}}, \ and\ \bibinfo {author}
  {\bibfnamefont {Y.}~\bibnamefont {Utsuno}},\ }\href {\doibase
  10.1103/RevModPhys.92.015002} {\bibfield  {journal} {\bibinfo  {journal}
  {Rev. Mod. Phys.}\ }\textbf {\bibinfo {volume} {92}},\ \bibinfo {pages}
  {015002} (\bibinfo {year} {2020})}\BibitemShut {NoStop}%
\bibitem [{\citenamefont {Kajino}\ and\ \citenamefont
  {Mathews}(2017)}]{Kajino2017}%
  \BibitemOpen
  \bibfield  {author} {\bibinfo {author} {\bibfnamefont {T.}~\bibnamefont
  {Kajino}}\ and\ \bibinfo {author} {\bibfnamefont {G.~J.}\ \bibnamefont
  {Mathews}},\ }\href {\doibase 10.1088/1361-6633/aa6a25} {\bibfield  {journal}
  {\bibinfo  {journal} {Reports on Progress in Physics}\ }\textbf {\bibinfo
  {volume} {80}},\ \bibinfo {pages} {084901} (\bibinfo {year}
  {2017})}\BibitemShut {NoStop}%
\bibitem [{\citenamefont {Lorusso}\ \emph {et~al.}(2015)\citenamefont
  {Lorusso}, \citenamefont {Nishimura}, \citenamefont {Xu}, \citenamefont
  {Jungclaus}, \citenamefont {Shimizu}, \citenamefont {Simpson}, \citenamefont
  {S\"oderstr\"om}, \citenamefont {Watanabe}, \citenamefont {Browne},
  \citenamefont {Doornenbal}, \citenamefont {Gey}, \citenamefont {Jung},
  \citenamefont {Meyer}, \citenamefont {Sumikama}, \citenamefont {Taprogge},
  \citenamefont {Vajta}, \citenamefont {Wu}, \citenamefont {Baba},
  \citenamefont {Benzoni}, \citenamefont {Chae}, \citenamefont {Crespi},
  \citenamefont {Fukuda}, \citenamefont {Gernh\"auser}, \citenamefont {Inabe},
  \citenamefont {Isobe}, \citenamefont {Kajino}, \citenamefont {Kameda},
  \citenamefont {Kim}, \citenamefont {Kim}, \citenamefont {Kojouharov},
  \citenamefont {Kondev}, \citenamefont {Kubo}, \citenamefont {Kurz},
  \citenamefont {Kwon}, \citenamefont {Lane}, \citenamefont {Li}, \citenamefont
  {Montaner-Piz\'a}, \citenamefont {Moschner}, \citenamefont {Naqvi},
  \citenamefont {Niikura}, \citenamefont {Nishibata}, \citenamefont {Odahara},
  \citenamefont {Orlandi}, \citenamefont {Patel}, \citenamefont {Podoly\'ak},
  \citenamefont {Sakurai}, \citenamefont {Schaffner}, \citenamefont {Schury},
  \citenamefont {Shibagaki}, \citenamefont {Steiger}, \citenamefont {Suzuki},
  \citenamefont {Takeda}, \citenamefont {Wendt}, \citenamefont {Yagi},\ and\
  \citenamefont {Yoshinaga}}]{Lorusso2015}%
  \BibitemOpen
  \bibfield  {author} {\bibinfo {author} {\bibfnamefont {G.}~\bibnamefont
  {Lorusso}}, \bibinfo {author} {\bibfnamefont {S.}~\bibnamefont {Nishimura}},
  \bibinfo {author} {\bibfnamefont {Z.~Y.}\ \bibnamefont {Xu}}, \bibinfo
  {author} {\bibfnamefont {A.}~\bibnamefont {Jungclaus}}, \bibinfo {author}
  {\bibfnamefont {Y.}~\bibnamefont {Shimizu}}, \bibinfo {author} {\bibfnamefont
  {G.~S.}\ \bibnamefont {Simpson}}, \bibinfo {author} {\bibfnamefont {P.-A.}\
  \bibnamefont {S\"oderstr\"om}}, \bibinfo {author} {\bibfnamefont
  {H.}~\bibnamefont {Watanabe}}, \bibinfo {author} {\bibfnamefont
  {F.}~\bibnamefont {Browne}}, \bibinfo {author} {\bibfnamefont
  {P.}~\bibnamefont {Doornenbal}}, \bibinfo {author} {\bibfnamefont
  {G.}~\bibnamefont {Gey}}, \bibinfo {author} {\bibfnamefont {H.~S.}\
  \bibnamefont {Jung}}, \bibinfo {author} {\bibfnamefont {B.}~\bibnamefont
  {Meyer}}, \bibinfo {author} {\bibfnamefont {T.}~\bibnamefont {Sumikama}},
  \bibinfo {author} {\bibfnamefont {J.}~\bibnamefont {Taprogge}}, \bibinfo
  {author} {\bibfnamefont {Z.}~\bibnamefont {Vajta}}, \bibinfo {author}
  {\bibfnamefont {J.}~\bibnamefont {Wu}}, \bibinfo {author} {\bibfnamefont
  {H.}~\bibnamefont {Baba}}, \bibinfo {author} {\bibfnamefont {G.}~\bibnamefont
  {Benzoni}}, \bibinfo {author} {\bibfnamefont {K.~Y.}\ \bibnamefont {Chae}},
  \bibinfo {author} {\bibfnamefont {F.~C.~L.}\ \bibnamefont {Crespi}}, \bibinfo
  {author} {\bibfnamefont {N.}~\bibnamefont {Fukuda}}, \bibinfo {author}
  {\bibfnamefont {R.}~\bibnamefont {Gernh\"auser}}, \bibinfo {author}
  {\bibfnamefont {N.}~\bibnamefont {Inabe}}, \bibinfo {author} {\bibfnamefont
  {T.}~\bibnamefont {Isobe}}, \bibinfo {author} {\bibfnamefont
  {T.}~\bibnamefont {Kajino}}, \bibinfo {author} {\bibfnamefont
  {D.}~\bibnamefont {Kameda}}, \bibinfo {author} {\bibfnamefont {G.~D.}\
  \bibnamefont {Kim}}, \bibinfo {author} {\bibfnamefont {Y.-K.}\ \bibnamefont
  {Kim}}, \bibinfo {author} {\bibfnamefont {I.}~\bibnamefont {Kojouharov}},
  \bibinfo {author} {\bibfnamefont {F.~G.}\ \bibnamefont {Kondev}}, \bibinfo
  {author} {\bibfnamefont {T.}~\bibnamefont {Kubo}}, \bibinfo {author}
  {\bibfnamefont {N.}~\bibnamefont {Kurz}}, \bibinfo {author} {\bibfnamefont
  {Y.~K.}\ \bibnamefont {Kwon}}, \bibinfo {author} {\bibfnamefont {G.~J.}\
  \bibnamefont {Lane}}, \bibinfo {author} {\bibfnamefont {Z.}~\bibnamefont
  {Li}}, \bibinfo {author} {\bibfnamefont {A.}~\bibnamefont {Montaner-Piz\'a}},
  \bibinfo {author} {\bibfnamefont {K.}~\bibnamefont {Moschner}}, \bibinfo
  {author} {\bibfnamefont {F.}~\bibnamefont {Naqvi}}, \bibinfo {author}
  {\bibfnamefont {M.}~\bibnamefont {Niikura}}, \bibinfo {author} {\bibfnamefont
  {H.}~\bibnamefont {Nishibata}}, \bibinfo {author} {\bibfnamefont
  {A.}~\bibnamefont {Odahara}}, \bibinfo {author} {\bibfnamefont
  {R.}~\bibnamefont {Orlandi}}, \bibinfo {author} {\bibfnamefont
  {Z.}~\bibnamefont {Patel}}, \bibinfo {author} {\bibfnamefont
  {Z.}~\bibnamefont {Podoly\'ak}}, \bibinfo {author} {\bibfnamefont
  {H.}~\bibnamefont {Sakurai}}, \bibinfo {author} {\bibfnamefont
  {H.}~\bibnamefont {Schaffner}}, \bibinfo {author} {\bibfnamefont
  {P.}~\bibnamefont {Schury}}, \bibinfo {author} {\bibfnamefont
  {S.}~\bibnamefont {Shibagaki}}, \bibinfo {author} {\bibfnamefont
  {K.}~\bibnamefont {Steiger}}, \bibinfo {author} {\bibfnamefont
  {H.}~\bibnamefont {Suzuki}}, \bibinfo {author} {\bibfnamefont
  {H.}~\bibnamefont {Takeda}}, \bibinfo {author} {\bibfnamefont
  {A.}~\bibnamefont {Wendt}}, \bibinfo {author} {\bibfnamefont
  {A.}~\bibnamefont {Yagi}}, \ and\ \bibinfo {author} {\bibfnamefont
  {K.}~\bibnamefont {Yoshinaga}},\ }\href {\doibase
  10.1103/PhysRevLett.114.192501} {\bibfield  {journal} {\bibinfo  {journal}
  {Phys. Rev. Lett.}\ }\textbf {\bibinfo {volume} {114}},\ \bibinfo {pages}
  {192501} (\bibinfo {year} {2015})}\BibitemShut {NoStop}%
\bibitem [{\citenamefont {Wu}\ \emph {et~al.}(2020)\citenamefont {Wu},
  \citenamefont {Nishimura}, \citenamefont {M\"oller}, \citenamefont
  {Mumpower}, \citenamefont {Lozeva}, \citenamefont {Moon}, \citenamefont
  {Odahara}, \citenamefont {Baba}, \citenamefont {Browne}, \citenamefont
  {Daido}, \citenamefont {Doornenbal}, \citenamefont {Fang}, \citenamefont
  {Haroon}, \citenamefont {Isobe}, \citenamefont {Jung}, \citenamefont
  {Lorusso}, \citenamefont {Moon}, \citenamefont {Patel}, \citenamefont {Rice},
  \citenamefont {Sakurai}, \citenamefont {Shimizu}, \citenamefont {Sinclair},
  \citenamefont {S\"oderstr\"om}, \citenamefont {Sumikama}, \citenamefont
  {Watanabe}, \citenamefont {Xu}, \citenamefont {Yagi}, \citenamefont
  {Yokoyama}, \citenamefont {Ahn}, \citenamefont {Bello~Garrote}, \citenamefont
  {Daugas}, \citenamefont {Didierjean}, \citenamefont {Fukuda}, \citenamefont
  {Inabe}, \citenamefont {Ishigaki}, \citenamefont {Kameda}, \citenamefont
  {Kojouharov}, \citenamefont {Komatsubara}, \citenamefont {Kubo},
  \citenamefont {Kurz}, \citenamefont {Kwon}, \citenamefont {Morimoto},
  \citenamefont {Murai}, \citenamefont {Nishibata}, \citenamefont {Schaffner},
  \citenamefont {Sprouse}, \citenamefont {Suzuki}, \citenamefont {Takeda},
  \citenamefont {Tanaka}, \citenamefont {Tshoo},\ and\ \citenamefont
  {Wakabayashi}}]{RIKEN2020}%
  \BibitemOpen
  \bibfield  {author} {\bibinfo {author} {\bibfnamefont {J.}~\bibnamefont
  {Wu}}, \bibinfo {author} {\bibfnamefont {S.}~\bibnamefont {Nishimura}},
  \bibinfo {author} {\bibfnamefont {P.}~\bibnamefont {M\"oller}}, \bibinfo
  {author} {\bibfnamefont {M.~R.}\ \bibnamefont {Mumpower}}, \bibinfo {author}
  {\bibfnamefont {R.}~\bibnamefont {Lozeva}}, \bibinfo {author} {\bibfnamefont
  {C.~B.}\ \bibnamefont {Moon}}, \bibinfo {author} {\bibfnamefont
  {A.}~\bibnamefont {Odahara}}, \bibinfo {author} {\bibfnamefont
  {H.}~\bibnamefont {Baba}}, \bibinfo {author} {\bibfnamefont {F.}~\bibnamefont
  {Browne}}, \bibinfo {author} {\bibfnamefont {R.}~\bibnamefont {Daido}},
  \bibinfo {author} {\bibfnamefont {P.}~\bibnamefont {Doornenbal}}, \bibinfo
  {author} {\bibfnamefont {Y.~F.}\ \bibnamefont {Fang}}, \bibinfo {author}
  {\bibfnamefont {M.}~\bibnamefont {Haroon}}, \bibinfo {author} {\bibfnamefont
  {T.}~\bibnamefont {Isobe}}, \bibinfo {author} {\bibfnamefont {H.~S.}\
  \bibnamefont {Jung}}, \bibinfo {author} {\bibfnamefont {G.}~\bibnamefont
  {Lorusso}}, \bibinfo {author} {\bibfnamefont {B.}~\bibnamefont {Moon}},
  \bibinfo {author} {\bibfnamefont {Z.}~\bibnamefont {Patel}}, \bibinfo
  {author} {\bibfnamefont {S.}~\bibnamefont {Rice}}, \bibinfo {author}
  {\bibfnamefont {H.}~\bibnamefont {Sakurai}}, \bibinfo {author} {\bibfnamefont
  {Y.}~\bibnamefont {Shimizu}}, \bibinfo {author} {\bibfnamefont
  {L.}~\bibnamefont {Sinclair}}, \bibinfo {author} {\bibfnamefont {P.-A.}\
  \bibnamefont {S\"oderstr\"om}}, \bibinfo {author} {\bibfnamefont
  {T.}~\bibnamefont {Sumikama}}, \bibinfo {author} {\bibfnamefont
  {H.}~\bibnamefont {Watanabe}}, \bibinfo {author} {\bibfnamefont {Z.~Y.}\
  \bibnamefont {Xu}}, \bibinfo {author} {\bibfnamefont {A.}~\bibnamefont
  {Yagi}}, \bibinfo {author} {\bibfnamefont {R.}~\bibnamefont {Yokoyama}},
  \bibinfo {author} {\bibfnamefont {D.~S.}\ \bibnamefont {Ahn}}, \bibinfo
  {author} {\bibfnamefont {F.~L.}\ \bibnamefont {Bello~Garrote}}, \bibinfo
  {author} {\bibfnamefont {J.~M.}\ \bibnamefont {Daugas}}, \bibinfo {author}
  {\bibfnamefont {F.}~\bibnamefont {Didierjean}}, \bibinfo {author}
  {\bibfnamefont {N.}~\bibnamefont {Fukuda}}, \bibinfo {author} {\bibfnamefont
  {N.}~\bibnamefont {Inabe}}, \bibinfo {author} {\bibfnamefont
  {T.}~\bibnamefont {Ishigaki}}, \bibinfo {author} {\bibfnamefont
  {D.}~\bibnamefont {Kameda}}, \bibinfo {author} {\bibfnamefont
  {I.}~\bibnamefont {Kojouharov}}, \bibinfo {author} {\bibfnamefont
  {T.}~\bibnamefont {Komatsubara}}, \bibinfo {author} {\bibfnamefont
  {T.}~\bibnamefont {Kubo}}, \bibinfo {author} {\bibfnamefont {N.}~\bibnamefont
  {Kurz}}, \bibinfo {author} {\bibfnamefont {K.~Y.}\ \bibnamefont {Kwon}},
  \bibinfo {author} {\bibfnamefont {S.}~\bibnamefont {Morimoto}}, \bibinfo
  {author} {\bibfnamefont {D.}~\bibnamefont {Murai}}, \bibinfo {author}
  {\bibfnamefont {H.}~\bibnamefont {Nishibata}}, \bibinfo {author}
  {\bibfnamefont {H.}~\bibnamefont {Schaffner}}, \bibinfo {author}
  {\bibfnamefont {T.~M.}\ \bibnamefont {Sprouse}}, \bibinfo {author}
  {\bibfnamefont {H.}~\bibnamefont {Suzuki}}, \bibinfo {author} {\bibfnamefont
  {H.}~\bibnamefont {Takeda}}, \bibinfo {author} {\bibfnamefont
  {M.}~\bibnamefont {Tanaka}}, \bibinfo {author} {\bibfnamefont
  {K.}~\bibnamefont {Tshoo}}, \ and\ \bibinfo {author} {\bibfnamefont
  {Y.}~\bibnamefont {Wakabayashi}},\ }\href {\doibase
  10.1103/PhysRevC.101.042801} {\bibfield  {journal} {\bibinfo  {journal}
  {Phys. Rev. C}\ }\textbf {\bibinfo {volume} {101}},\ \bibinfo {pages}
  {042801} (\bibinfo {year} {2020})}\BibitemShut {NoStop}%
\bibitem [{\citenamefont {Engel}\ \emph {et~al.}(1999)\citenamefont {Engel},
  \citenamefont {Bender}, \citenamefont {Dobaczewski}, \citenamefont
  {Nazarewicz},\ and\ \citenamefont {Surman}}]{Engel1999}%
  \BibitemOpen
  \bibfield  {author} {\bibinfo {author} {\bibfnamefont {J.}~\bibnamefont
  {Engel}}, \bibinfo {author} {\bibfnamefont {M.}~\bibnamefont {Bender}},
  \bibinfo {author} {\bibfnamefont {J.}~\bibnamefont {Dobaczewski}}, \bibinfo
  {author} {\bibfnamefont {W.}~\bibnamefont {Nazarewicz}}, \ and\ \bibinfo
  {author} {\bibfnamefont {R.}~\bibnamefont {Surman}},\ }\href@noop {}
  {\bibfield  {journal} {\bibinfo  {journal} {Physiscal Review C}\ }\textbf
  {\bibinfo {volume} {60}} (\bibinfo {year} {1999})}\BibitemShut {NoStop}%
\bibitem [{\citenamefont {Nik\ifmmode \check{s}\else
  \v{s}\fi{}i\ifmmode~\acute{c}\else \'{c}\fi{}}\ \emph
  {et~al.}(2005)\citenamefont {Nik\ifmmode \check{s}\else
  \v{s}\fi{}i\ifmmode~\acute{c}\else \'{c}\fi{}}, \citenamefont {Marketin},
  \citenamefont {Vretenar}, \citenamefont {Paar},\ and\ \citenamefont
  {Ring}}]{Niksic2005}%
  \BibitemOpen
  \bibfield  {author} {\bibinfo {author} {\bibfnamefont {T.}~\bibnamefont
  {Nik\ifmmode \check{s}\else \v{s}\fi{}i\ifmmode~\acute{c}\else \'{c}\fi{}}},
  \bibinfo {author} {\bibfnamefont {T.}~\bibnamefont {Marketin}}, \bibinfo
  {author} {\bibfnamefont {D.}~\bibnamefont {Vretenar}}, \bibinfo {author}
  {\bibfnamefont {N.}~\bibnamefont {Paar}}, \ and\ \bibinfo {author}
  {\bibfnamefont {P.}~\bibnamefont {Ring}},\ }\href {\doibase
  10.1103/PhysRevC.71.014308} {\bibfield  {journal} {\bibinfo  {journal} {Phys.
  Rev. C}\ }\textbf {\bibinfo {volume} {71}},\ \bibinfo {pages} {014308}
  (\bibinfo {year} {2005})}\BibitemShut {NoStop}%
\bibitem [{\citenamefont {Yoshida}(2013)}]{Yoshida2013}%
  \BibitemOpen
  \bibfield  {author} {\bibinfo {author} {\bibfnamefont {K.}~\bibnamefont
  {Yoshida}},\ }\href {\doibase 10.1093/ptep/ptt091} {\bibfield  {journal}
  {\bibinfo  {journal} {Progress of Theoretical and Experimental Physics}\
  }\textbf {\bibinfo {volume} {2013}} (\bibinfo {year} {2013}),\
  10.1093/ptep/ptt091},\ \bibinfo {note} {113D02},\ \Eprint
  {http://arxiv.org/abs/https://academic.oup.com/ptep/article-pdf/2013/11/113D02/9719090/ptt091.pdf}
  {https://academic.oup.com/ptep/article-pdf/2013/11/113D02/9719090/ptt091.pdf}
  \BibitemShut {NoStop}%
\bibitem [{\citenamefont {Niu}\ \emph {et~al.}(2013)\citenamefont {Niu},
  \citenamefont {Niu}, \citenamefont {Liang}, \citenamefont {Long},
  \citenamefont {Nik\v{s}ic}, \citenamefont {Vretenar},\ and\ \citenamefont
  {Meng}}]{Niu2013}%
  \BibitemOpen
  \bibfield  {author} {\bibinfo {author} {\bibfnamefont {Z.}~\bibnamefont
  {Niu}}, \bibinfo {author} {\bibfnamefont {Y.}~\bibnamefont {Niu}}, \bibinfo
  {author} {\bibfnamefont {H.}~\bibnamefont {Liang}}, \bibinfo {author}
  {\bibfnamefont {W.}~\bibnamefont {Long}}, \bibinfo {author} {\bibfnamefont
  {T.}~\bibnamefont {Nik\v{s}ic}}, \bibinfo {author} {\bibfnamefont
  {D.}~\bibnamefont {Vretenar}}, \ and\ \bibinfo {author} {\bibfnamefont
  {J.}~\bibnamefont {Meng}},\ }\href {\doibase
  https://doi.org/10.1016/j.physletb.2013.04.048} {\bibfield  {journal}
  {\bibinfo  {journal} {Physics Letters B}\ }\textbf {\bibinfo {volume}
  {723}},\ \bibinfo {pages} {172} (\bibinfo {year} {2013})}\BibitemShut
  {NoStop}%
\bibitem [{\citenamefont {Ni}\ and\ \citenamefont
  {Ren}(2014)}]{PhysRevC.89.064320}%
  \BibitemOpen
  \bibfield  {author} {\bibinfo {author} {\bibfnamefont {D.}~\bibnamefont
  {Ni}}\ and\ \bibinfo {author} {\bibfnamefont {Z.}~\bibnamefont {Ren}},\
  }\href {\doibase 10.1103/PhysRevC.89.064320} {\bibfield  {journal} {\bibinfo
  {journal} {Phys. Rev. C}\ }\textbf {\bibinfo {volume} {89}},\ \bibinfo
  {pages} {064320} (\bibinfo {year} {2014})}\BibitemShut {NoStop}%
\bibitem [{\citenamefont {Martini}\ \emph {et~al.}(2014)\citenamefont
  {Martini}, \citenamefont {P\'eru},\ and\ \citenamefont
  {Goriely}}]{Martini2014}%
  \BibitemOpen
  \bibfield  {author} {\bibinfo {author} {\bibfnamefont {M.}~\bibnamefont
  {Martini}}, \bibinfo {author} {\bibfnamefont {S.}~\bibnamefont {P\'eru}}, \
  and\ \bibinfo {author} {\bibfnamefont {S.}~\bibnamefont {Goriely}},\ }\href
  {\doibase 10.1103/PhysRevC.89.044306} {\bibfield  {journal} {\bibinfo
  {journal} {Phys. Rev. C}\ }\textbf {\bibinfo {volume} {89}},\ \bibinfo
  {pages} {044306} (\bibinfo {year} {2014})}\BibitemShut {NoStop}%
\bibitem [{\citenamefont {Sarriguren}(2015)}]{Sarriguren2015}%
  \BibitemOpen
  \bibfield  {author} {\bibinfo {author} {\bibfnamefont {P.}~\bibnamefont
  {Sarriguren}},\ }\href {\doibase 10.1103/PhysRevC.91.044304} {\bibfield
  {journal} {\bibinfo  {journal} {Phys. Rev. C}\ }\textbf {\bibinfo {volume}
  {91}},\ \bibinfo {pages} {044304} (\bibinfo {year} {2015})}\BibitemShut
  {NoStop}%
\bibitem [{\citenamefont {{Minato, Futoshi}}(2016)}]{Minato2016}%
  \BibitemOpen
  \bibfield  {author} {\bibinfo {author} {\bibnamefont {{Minato, Futoshi}}},\
  }\href {\doibase 10.1051/epjconf/201612210001} {\bibfield  {journal}
  {\bibinfo  {journal} {EPJ Web of Conferences}\ }\textbf {\bibinfo {volume}
  {122}},\ \bibinfo {pages} {10001} (\bibinfo {year} {2016})}\BibitemShut
  {NoStop}%
\bibitem [{\citenamefont {Wang}\ \emph {et~al.}(2016)\citenamefont {Wang},
  \citenamefont {Niu}, \citenamefont {Niu},\ and\ \citenamefont
  {Guo}}]{Wang2016}%
  \BibitemOpen
  \bibfield  {author} {\bibinfo {author} {\bibfnamefont {Z.~Y.}\ \bibnamefont
  {Wang}}, \bibinfo {author} {\bibfnamefont {Y.~F.}\ \bibnamefont {Niu}},
  \bibinfo {author} {\bibfnamefont {Z.~M.}\ \bibnamefont {Niu}}, \ and\
  \bibinfo {author} {\bibfnamefont {J.~Y.}\ \bibnamefont {Guo}},\ }\href
  {\doibase 10.1088/0954-3899/43/4/045108} {\bibfield  {journal} {\bibinfo
  {journal} {Journal of Physics G: Nuclear and Particle Physics}\ }\textbf
  {\bibinfo {volume} {43}},\ \bibinfo {pages} {045108} (\bibinfo {year}
  {2016})}\BibitemShut {NoStop}%
\bibitem [{\citenamefont {Mustonen}\ and\ \citenamefont
  {Engel}(2016)}]{Mustonen2016}%
  \BibitemOpen
  \bibfield  {author} {\bibinfo {author} {\bibfnamefont {M.~T.}\ \bibnamefont
  {Mustonen}}\ and\ \bibinfo {author} {\bibfnamefont {J.}~\bibnamefont
  {Engel}},\ }\href {\doibase 10.1103/PhysRevC.93.014304} {\bibfield  {journal}
  {\bibinfo  {journal} {Phys. Rev. C}\ }\textbf {\bibinfo {volume} {93}},\
  \bibinfo {pages} {014304} (\bibinfo {year} {2016})}\BibitemShut {NoStop}%
\bibitem [{\citenamefont {Borzov}(2020)}]{Borzov2020}%
  \BibitemOpen
  \bibfield  {author} {\bibinfo {author} {\bibfnamefont {I.~N.}\ \bibnamefont
  {Borzov}},\ }\href {\doibase 10.1134/S1063778820050087} {\bibfield  {journal}
  {\bibinfo  {journal} {Physics of Atomic Nuclei}\ }\textbf {\bibinfo {volume}
  {83}},\ \bibinfo {pages} {700} (\bibinfo {year} {2020})}\BibitemShut
  {NoStop}%
\bibitem [{\citenamefont {Wen}\ \emph {et~al.}(2021)\citenamefont {Wen},
  \citenamefont {Zhang}, \citenamefont {Cao},\ and\ \citenamefont
  {Zhang}}]{Wen2021}%
  \BibitemOpen
  \bibfield  {author} {\bibinfo {author} {\bibfnamefont {P.-W.}\ \bibnamefont
  {Wen}}, \bibinfo {author} {\bibfnamefont {S.-S.}\ \bibnamefont {Zhang}},
  \bibinfo {author} {\bibfnamefont {L.-G.}\ \bibnamefont {Cao}}, \ and\
  \bibinfo {author} {\bibfnamefont {F.-S.}\ \bibnamefont {Zhang}},\ }\href
  {\doibase 10.1088/1674-1137/abc1d1} {\bibfield  {journal} {\bibinfo
  {journal} {Chinese Physics C}\ }\textbf {\bibinfo {volume} {45}},\ \bibinfo
  {pages} {014105} (\bibinfo {year} {2021})}\BibitemShut {NoStop}%
\bibitem [{\citenamefont {Suzuki}\ \emph {et~al.}(2012)\citenamefont {Suzuki},
  \citenamefont {Yoshida}, \citenamefont {Kajino},\ and\ \citenamefont
  {Otsuka}}]{Suzuki2012}%
  \BibitemOpen
  \bibfield  {author} {\bibinfo {author} {\bibfnamefont {T.}~\bibnamefont
  {Suzuki}}, \bibinfo {author} {\bibfnamefont {T.}~\bibnamefont {Yoshida}},
  \bibinfo {author} {\bibfnamefont {T.}~\bibnamefont {Kajino}}, \ and\ \bibinfo
  {author} {\bibfnamefont {T.}~\bibnamefont {Otsuka}},\ }\href {\doibase
  10.1103/PhysRevC.85.015802} {\bibfield  {journal} {\bibinfo  {journal} {Phys.
  Rev. C}\ }\textbf {\bibinfo {volume} {85}},\ \bibinfo {pages} {015802}
  (\bibinfo {year} {2012})}\BibitemShut {NoStop}%
\bibitem [{\citenamefont {Suzuki}\ \emph {et~al.}(2018)\citenamefont {Suzuki},
  \citenamefont {Shibagaki}, \citenamefont {Yoshida}, \citenamefont {Kajino},\
  and\ \citenamefont {Otsuka}}]{Suzuki2018}%
  \BibitemOpen
  \bibfield  {author} {\bibinfo {author} {\bibfnamefont {T.}~\bibnamefont
  {Suzuki}}, \bibinfo {author} {\bibfnamefont {S.}~\bibnamefont {Shibagaki}},
  \bibinfo {author} {\bibfnamefont {T.}~\bibnamefont {Yoshida}}, \bibinfo
  {author} {\bibfnamefont {T.}~\bibnamefont {Kajino}}, \ and\ \bibinfo {author}
  {\bibfnamefont {T.}~\bibnamefont {Otsuka}},\ }\href {\doibase
  10.3847/1538-4357/aabfde} {\bibfield  {journal} {\bibinfo  {journal} {The
  Astrophysical Journal}\ }\textbf {\bibinfo {volume} {859}},\ \bibinfo {pages}
  {133} (\bibinfo {year} {2018})}\BibitemShut {NoStop}%
\bibitem [{\citenamefont {Suzuki}\ \emph {et~al.}(2019)\citenamefont {Suzuki},
  \citenamefont {Shibagaki}, \citenamefont {Yoshida}, \citenamefont {Kajino},\
  and\ \citenamefont {Otsuka}}]{Suzuki2019}%
  \BibitemOpen
  \bibfield  {author} {\bibinfo {author} {\bibfnamefont {T.}~\bibnamefont
  {Suzuki}}, \bibinfo {author} {\bibfnamefont {S.}~\bibnamefont {Shibagaki}},
  \bibinfo {author} {\bibfnamefont {T.}~\bibnamefont {Yoshida}}, \bibinfo
  {author} {\bibfnamefont {T.}~\bibnamefont {Kajino}}, \ and\ \bibinfo {author}
  {\bibfnamefont {T.}~\bibnamefont {Otsuka}},\ }\enquote {\bibinfo {title}
  {R-process nucleosynthesis in core-collapse supernova explosions and binary
  neutron star mergers.}}\ in\ \href {\doibase
  https://doi.org/10.1007/978-3-030-13876-9_84} {\emph {\bibinfo {booktitle}
  {Nuclei in the Cosmos XV., Springer Proceedings in Physics}}},\ Vol.\
  \bibinfo {volume} {219},\ \bibinfo {editor} {edited by\ \bibinfo {editor}
  {\bibfnamefont {A.}~\bibnamefont {Formicola}}, \bibinfo {editor}
  {\bibfnamefont {M.}~\bibnamefont {Junker}}, \bibinfo {editor} {\bibfnamefont
  {L.}~\bibnamefont {Gialanella}}, \ and\ \bibinfo {editor} {\bibfnamefont
  {G.}~\bibnamefont {Imbriani}}}\ (\bibinfo  {publisher} {Springer, Cham},\
  \bibinfo {year} {2019})\ p.\ \bibinfo {pages} {437}\BibitemShut {NoStop}%
\bibitem [{\citenamefont {Nomura}\ \emph {et~al.}(2020)\citenamefont {Nomura},
  \citenamefont {Rodr\'{\i}guez-Guzm\'an},\ and\ \citenamefont
  {Robledo}}]{Nomura2020}%
  \BibitemOpen
  \bibfield  {author} {\bibinfo {author} {\bibfnamefont {K.}~\bibnamefont
  {Nomura}}, \bibinfo {author} {\bibfnamefont {R.}~\bibnamefont
  {Rodr\'{\i}guez-Guzm\'an}}, \ and\ \bibinfo {author} {\bibfnamefont {L.~M.}\
  \bibnamefont {Robledo}},\ }\href {\doibase 10.1103/PhysRevC.101.044318}
  {\bibfield  {journal} {\bibinfo  {journal} {Phys. Rev. C}\ }\textbf {\bibinfo
  {volume} {101}},\ \bibinfo {pages} {044318} (\bibinfo {year}
  {2020})}\BibitemShut {NoStop}%
\bibitem [{\citenamefont {Marketin}\ \emph {et~al.}(2016)\citenamefont
  {Marketin}, \citenamefont {Huther},\ and\ \citenamefont
  {Martinez-Pinedo}}]{Tomislav2016}%
  \BibitemOpen
  \bibfield  {author} {\bibinfo {author} {\bibfnamefont {T.}~\bibnamefont
  {Marketin}}, \bibinfo {author} {\bibfnamefont {L.}~\bibnamefont {Huther}}, \
  and\ \bibinfo {author} {\bibfnamefont {G.}~\bibnamefont {Martinez-Pinedo}},\
  }\href@noop {} {\bibfield  {journal} {\bibinfo  {journal} {Phys. Rev. C}\
  }\textbf {\bibinfo {volume} {93}} (\bibinfo {year} {2016})}\BibitemShut
  {NoStop}%
\bibitem [{\citenamefont {Ney}\ \emph {et~al.}(2020)\citenamefont {Ney},
  \citenamefont {Engel}, \citenamefont {Li},\ and\ \citenamefont
  {Schunck}}]{Ney2020}%
  \BibitemOpen
  \bibfield  {author} {\bibinfo {author} {\bibfnamefont {E.~M.}\ \bibnamefont
  {Ney}}, \bibinfo {author} {\bibfnamefont {J.}~\bibnamefont {Engel}}, \bibinfo
  {author} {\bibfnamefont {T.}~\bibnamefont {Li}}, \ and\ \bibinfo {author}
  {\bibfnamefont {N.}~\bibnamefont {Schunck}},\ }\href {\doibase
  10.1103/PhysRevC.102.034326} {\bibfield  {journal} {\bibinfo  {journal}
  {Phys. Rev. C}\ }\textbf {\bibinfo {volume} {102}},\ \bibinfo {pages}
  {034326} (\bibinfo {year} {2020})}\BibitemShut {NoStop}%
\bibitem [{\citenamefont {Marketin}\ \emph {et~al.}(2007)\citenamefont
  {Marketin}, \citenamefont {Vretenar},\ and\ \citenamefont
  {Ring}}]{Marketin2007}%
  \BibitemOpen
  \bibfield  {author} {\bibinfo {author} {\bibfnamefont {T.}~\bibnamefont
  {Marketin}}, \bibinfo {author} {\bibfnamefont {D.}~\bibnamefont {Vretenar}},
  \ and\ \bibinfo {author} {\bibfnamefont {P.}~\bibnamefont {Ring}},\ }\href
  {\doibase 10.1103/PhysRevC.75.024304} {\bibfield  {journal} {\bibinfo
  {journal} {Phys. Rev. C}\ }\textbf {\bibinfo {volume} {75}},\ \bibinfo
  {pages} {024304} (\bibinfo {year} {2007})}\BibitemShut {NoStop}%
\bibitem [{\citenamefont {Takahara}\ \emph {et~al.}(1994)\citenamefont
  {Takahara}, \citenamefont {Onishi},\ and\ \citenamefont
  {Tajima}}]{Takahara1994}%
  \BibitemOpen
  \bibfield  {author} {\bibinfo {author} {\bibfnamefont {S.}~\bibnamefont
  {Takahara}}, \bibinfo {author} {\bibfnamefont {N.}~\bibnamefont {Onishi}}, \
  and\ \bibinfo {author} {\bibfnamefont {N.}~\bibnamefont {Tajima}},\ }\href
  {\doibase https://doi.org/10.1016/0370-2693(94)91048-0} {\bibfield  {journal}
  {\bibinfo  {journal} {Physics Letters B}\ }\textbf {\bibinfo {volume}
  {331}},\ \bibinfo {pages} {261} (\bibinfo {year} {1994})}\BibitemShut
  {NoStop}%
\bibitem [{\citenamefont {Utama}\ \emph {et~al.}(2016)\citenamefont {Utama},
  \citenamefont {Piekarewicz},\ and\ \citenamefont
  {Prosper}}]{Utama2016Phys.Rev.C93_014311}%
  \BibitemOpen
  \bibfield  {author} {\bibinfo {author} {\bibfnamefont {R.}~\bibnamefont
  {Utama}}, \bibinfo {author} {\bibfnamefont {J.}~\bibnamefont {Piekarewicz}},
  \ and\ \bibinfo {author} {\bibfnamefont {H.~B.}\ \bibnamefont {Prosper}},\
  }\href {\doibase 10.1103/PhysRevC.93.014311} {\bibfield  {journal} {\bibinfo
  {journal} {Phys. Rev. C}\ }\textbf {\bibinfo {volume} {93}},\ \bibinfo
  {pages} {014311} (\bibinfo {year} {2016})}\BibitemShut {NoStop}%
\bibitem [{\citenamefont {Niu}\ and\ \citenamefont {Liang}(2018)}]{ZM2018}%
  \BibitemOpen
  \bibfield  {author} {\bibinfo {author} {\bibfnamefont {Z.}~\bibnamefont
  {Niu}}\ and\ \bibinfo {author} {\bibfnamefont {H.}~\bibnamefont {Liang}},\
  }\href {\doibase https://doi.org/10.1016/j.physletb.2018.01.002} {\bibfield
  {journal} {\bibinfo  {journal} {Physics Letters B}\ }\textbf {\bibinfo
  {volume} {778}},\ \bibinfo {pages} {48} (\bibinfo {year} {2018})}\BibitemShut
  {NoStop}%
\bibitem [{\citenamefont {Niu}\ \emph {et~al.}(2019)\citenamefont {Niu},
  \citenamefont {Liang}, \citenamefont {Sun}, \citenamefont {Long},\ and\
  \citenamefont {Niu}}]{ZM2019}%
  \BibitemOpen
  \bibfield  {author} {\bibinfo {author} {\bibfnamefont {Z.~M.}\ \bibnamefont
  {Niu}}, \bibinfo {author} {\bibfnamefont {H.~Z.}\ \bibnamefont {Liang}},
  \bibinfo {author} {\bibfnamefont {B.~H.}\ \bibnamefont {Sun}}, \bibinfo
  {author} {\bibfnamefont {W.~H.}\ \bibnamefont {Long}}, \ and\ \bibinfo
  {author} {\bibfnamefont {Y.~F.}\ \bibnamefont {Niu}},\ }\href {\doibase
  10.1103/PhysRevC.99.064307} {\bibfield  {journal} {\bibinfo  {journal} {Phys.
  Rev. C}\ }\textbf {\bibinfo {volume} {99}},\ \bibinfo {pages} {064307}
  (\bibinfo {year} {2019})}\BibitemShut {NoStop}%
\bibitem [{\citenamefont {Reinhard}\ \emph {et~al.}(1999)\citenamefont
  {Reinhard}, \citenamefont {Dean}, \citenamefont {Nazarewicz}, \citenamefont
  {Dobaczewski}, \citenamefont {Maruhn},\ and\ \citenamefont
  {Strayer}}]{Reinhard1999}%
  \BibitemOpen
  \bibfield  {author} {\bibinfo {author} {\bibfnamefont {P.-G.}\ \bibnamefont
  {Reinhard}}, \bibinfo {author} {\bibfnamefont {D.~J.}\ \bibnamefont {Dean}},
  \bibinfo {author} {\bibfnamefont {W.}~\bibnamefont {Nazarewicz}}, \bibinfo
  {author} {\bibfnamefont {J.}~\bibnamefont {Dobaczewski}}, \bibinfo {author}
  {\bibfnamefont {J.~A.}\ \bibnamefont {Maruhn}}, \ and\ \bibinfo {author}
  {\bibfnamefont {M.~R.}\ \bibnamefont {Strayer}},\ }\href {\doibase
  10.1103/PhysRevC.60.014316} {\bibfield  {journal} {\bibinfo  {journal} {Phys.
  Rev. C}\ }\textbf {\bibinfo {volume} {60}},\ \bibinfo {pages} {014316}
  (\bibinfo {year} {1999})}\BibitemShut {NoStop}%
\bibitem [{\citenamefont {Berger}\ \emph {et~al.}(1984)\citenamefont {Berger},
  \citenamefont {Girod},\ and\ \citenamefont {Gogny}}]{Berger1984}%
  \BibitemOpen
  \bibfield  {author} {\bibinfo {author} {\bibfnamefont {J.}~\bibnamefont
  {Berger}}, \bibinfo {author} {\bibfnamefont {M.}~\bibnamefont {Girod}}, \
  and\ \bibinfo {author} {\bibfnamefont {D.}~\bibnamefont {Gogny}},\ }\href
  {\doibase https://doi.org/10.1016/0375-9474(84)90240-9} {\bibfield  {journal}
  {\bibinfo  {journal} {Nuclear Physics A}\ }\textbf {\bibinfo {volume}
  {428}},\ \bibinfo {pages} {23 } (\bibinfo {year} {1984})}\BibitemShut
  {NoStop}%
\bibitem [{\citenamefont {Garrido}\ \emph {et~al.}(1999)\citenamefont
  {Garrido}, \citenamefont {Sarriguren}, \citenamefont {Moya~de Guerra},\ and\
  \citenamefont {Schuck}}]{Garrido1999}%
  \BibitemOpen
  \bibfield  {author} {\bibinfo {author} {\bibfnamefont {E.}~\bibnamefont
  {Garrido}}, \bibinfo {author} {\bibfnamefont {P.}~\bibnamefont {Sarriguren}},
  \bibinfo {author} {\bibfnamefont {E.}~\bibnamefont {Moya~de Guerra}}, \ and\
  \bibinfo {author} {\bibfnamefont {P.}~\bibnamefont {Schuck}},\ }\href
  {\doibase 10.1103/PhysRevC.60.064312} {\bibfield  {journal} {\bibinfo
  {journal} {Phys. Rev. C}\ }\textbf {\bibinfo {volume} {60}},\ \bibinfo
  {pages} {064312} (\bibinfo {year} {1999})}\BibitemShut {NoStop}%
\bibitem [{\citenamefont {Satu{\l}a}(2006)}]{Satula2006}%
  \BibitemOpen
  \bibfield  {author} {\bibinfo {author} {\bibfnamefont {W.}~\bibnamefont
  {Satu{\l}a}},\ }\href {\doibase 10.1088/0031-8949/2006/t125/018} {\bibfield
  {journal} {\bibinfo  {journal} {Physica Scripta}\ }\textbf {\bibinfo {volume}
  {T125}},\ \bibinfo {pages} {82} (\bibinfo {year} {2006})}\BibitemShut
  {NoStop}%
\bibitem [{\citenamefont {Dobaczewski}\ \emph {et~al.}(1984)\citenamefont
  {Dobaczewski}, \citenamefont {Flocard},\ and\ \citenamefont
  {Treiner}}]{Dobaczewski1984}%
  \BibitemOpen
  \bibfield  {author} {\bibinfo {author} {\bibfnamefont {J.}~\bibnamefont
  {Dobaczewski}}, \bibinfo {author} {\bibfnamefont {H.}~\bibnamefont
  {Flocard}}, \ and\ \bibinfo {author} {\bibfnamefont {J.}~\bibnamefont
  {Treiner}},\ }\href {\doibase https://doi.org/10.1016/0375-9474(84)90433-0}
  {\bibfield  {journal} {\bibinfo  {journal} {Nuclear Physics A}\ }\textbf
  {\bibinfo {volume} {422}},\ \bibinfo {pages} {103} (\bibinfo {year}
  {1984})}\BibitemShut {NoStop}%
\bibitem [{\citenamefont {Stoitsov}\ \emph {et~al.}(2003)\citenamefont
  {Stoitsov}, \citenamefont {Dobaczewski}, \citenamefont {Nazarewicz},
  \citenamefont {Pittel},\ and\ \citenamefont {Dean}}]{Stoitsov2003}%
  \BibitemOpen
  \bibfield  {author} {\bibinfo {author} {\bibfnamefont {M.~V.}\ \bibnamefont
  {Stoitsov}}, \bibinfo {author} {\bibfnamefont {J.}~\bibnamefont
  {Dobaczewski}}, \bibinfo {author} {\bibfnamefont {W.}~\bibnamefont
  {Nazarewicz}}, \bibinfo {author} {\bibfnamefont {S.}~\bibnamefont {Pittel}},
  \ and\ \bibinfo {author} {\bibfnamefont {D.~J.}\ \bibnamefont {Dean}},\
  }\href {\doibase 10.1103/PhysRevC.68.054312} {\bibfield  {journal} {\bibinfo
  {journal} {Phys. Rev. C}\ }\textbf {\bibinfo {volume} {68}},\ \bibinfo
  {pages} {054312} (\bibinfo {year} {2003})}\BibitemShut {NoStop}%
\bibitem [{\citenamefont {Engel}\ \emph {et~al.}(1988)\citenamefont {Engel},
  \citenamefont {Vogel},\ and\ \citenamefont {Zirnbauer}}]{Engel1988}%
  \BibitemOpen
  \bibfield  {author} {\bibinfo {author} {\bibfnamefont {J.}~\bibnamefont
  {Engel}}, \bibinfo {author} {\bibfnamefont {P.}~\bibnamefont {Vogel}}, \ and\
  \bibinfo {author} {\bibfnamefont {M.~R.}\ \bibnamefont {Zirnbauer}},\ }\href
  {\doibase 10.1103/PhysRevC.37.731} {\bibfield  {journal} {\bibinfo  {journal}
  {Phys. Rev. C}\ }\textbf {\bibinfo {volume} {37}},\ \bibinfo {pages} {731}
  (\bibinfo {year} {1988})}\BibitemShut {NoStop}%
\bibitem [{\citenamefont {Behrens}\ and\ \citenamefont
  {B\"uhring}(1982)}]{Behrens}%
  \BibitemOpen
  \bibfield  {author} {\bibinfo {author} {\bibfnamefont {H.}~\bibnamefont
  {Behrens}}\ and\ \bibinfo {author} {\bibfnamefont {W.}~\bibnamefont
  {B\"uhring}},\ }\href@noop {} {\emph {\bibinfo {title} {Electron Radial Wave
  Functions and Nuclear $\beta$ Decay}}}\ (\bibinfo  {publisher} {Clarendon,
  Oxford},\ \bibinfo {year} {1982})\BibitemShut {NoStop}%
\bibitem [{\citenamefont {Hardy}\ and\ \citenamefont
  {Towner}(2009)}]{Hardy2009}%
  \BibitemOpen
  \bibfield  {author} {\bibinfo {author} {\bibfnamefont {J.~C.}\ \bibnamefont
  {Hardy}}\ and\ \bibinfo {author} {\bibfnamefont {I.~S.}\ \bibnamefont
  {Towner}},\ }\href {\doibase 10.1103/PhysRevC.79.055502} {\bibfield
  {journal} {\bibinfo  {journal} {Phys. Rev. C}\ }\textbf {\bibinfo {volume}
  {79}},\ \bibinfo {pages} {055502} (\bibinfo {year} {2009})}\BibitemShut
  {NoStop}%
\bibitem [{\citenamefont {Ring}\ and\ \citenamefont
  {Schuck}(1980)}]{RingandSchuck}%
  \BibitemOpen
  \bibfield  {author} {\bibinfo {author} {\bibfnamefont {P.}~\bibnamefont
  {Ring}}\ and\ \bibinfo {author} {\bibfnamefont {P.}~\bibnamefont {Schuck}},\
  }\href@noop {} {\emph {\bibinfo {title} {The Nuclear Many-Body Problem}}}\
  (\bibinfo  {publisher} {Springer-Verlag, Berlin},\ \bibinfo {year}
  {1980})\BibitemShut {NoStop}%
\bibitem [{\citenamefont {Minato}\ \emph {et~al.}(2021)\citenamefont {Minato},
  \citenamefont {Marketin},\ and\ \citenamefont {Paar}}]{Minato2021}%
  \BibitemOpen
  \bibfield  {author} {\bibinfo {author} {\bibfnamefont {F.}~\bibnamefont
  {Minato}}, \bibinfo {author} {\bibfnamefont {T.}~\bibnamefont {Marketin}}, \
  and\ \bibinfo {author} {\bibfnamefont {N.}~\bibnamefont {Paar}},\ }\href
  {\doibase 10.1103/PhysRevC.104.044321} {\bibfield  {journal} {\bibinfo
  {journal} {Phys. Rev. C}\ }\textbf {\bibinfo {volume} {104}},\ \bibinfo
  {pages} {044321} (\bibinfo {year} {2021})}\BibitemShut {NoStop}%
\bibitem [{\citenamefont {Behrens}\ and\ \citenamefont
  {B\"{u}hring}(1971)}]{Behrens1971}%
  \BibitemOpen
  \bibfield  {author} {\bibinfo {author} {\bibfnamefont {H.}~\bibnamefont
  {Behrens}}\ and\ \bibinfo {author} {\bibfnamefont {W.}~\bibnamefont
  {B\"{u}hring}},\ }\href {\doibase
  https://doi.org/10.1016/0375-9474(71)90489-1} {\bibfield  {journal} {\bibinfo
   {journal} {Nuclear Physics A}\ }\textbf {\bibinfo {volume} {162}},\ \bibinfo
  {pages} {111} (\bibinfo {year} {1971})}\BibitemShut {NoStop}%
\bibitem [{\citenamefont {{Particle Data Group and Zyla, P A}}\ \emph
  {et~al.}(2020)\citenamefont {{Particle Data Group and Zyla, P A}} \emph
  {et~al.}}]{PDG2020}%
  \BibitemOpen
  \bibfield  {author} {\bibinfo {author} {\bibnamefont {{Particle Data Group
  and Zyla, P A}}} \emph {et~al.},\ }\href@noop {} {\bibfield  {journal}
  {\bibinfo  {journal} {Progress of Theoretical and Experimental Physics}\
  }\textbf {\bibinfo {volume} {2020}} (\bibinfo {year} {2020})}\BibitemShut
  {NoStop}%
\bibitem [{\citenamefont {Men\'endez}\ \emph {et~al.}(2011)\citenamefont
  {Men\'endez}, \citenamefont {Gazit},\ and\ \citenamefont
  {Schwenk}}]{Menendez2011}%
  \BibitemOpen
  \bibfield  {author} {\bibinfo {author} {\bibfnamefont {J.}~\bibnamefont
  {Men\'endez}}, \bibinfo {author} {\bibfnamefont {D.}~\bibnamefont {Gazit}}, \
  and\ \bibinfo {author} {\bibfnamefont {A.}~\bibnamefont {Schwenk}},\ }\href
  {\doibase 10.1103/PhysRevLett.107.062501} {\bibfield  {journal} {\bibinfo
  {journal} {Phys. Rev. Lett.}\ }\textbf {\bibinfo {volume} {107}},\ \bibinfo
  {pages} {062501} (\bibinfo {year} {2011})}\BibitemShut {NoStop}%
\bibitem [{\citenamefont {Mart\'{i}nez-Pinedo}\ and\ \citenamefont
  {Poves}(1993)}]{Martinez-Pinedo1993}%
  \BibitemOpen
  \bibfield  {author} {\bibinfo {author} {\bibfnamefont {G.}~\bibnamefont
  {Mart\'{i}nez-Pinedo}}\ and\ \bibinfo {author} {\bibfnamefont
  {A.}~\bibnamefont {Poves}},\ }\href {\doibase 10.1103/PhysRevC.48.937}
  {\bibfield  {journal} {\bibinfo  {journal} {Phys. Rev. C}\ }\textbf {\bibinfo
  {volume} {48}},\ \bibinfo {pages} {937} (\bibinfo {year} {1993})}\BibitemShut
  {NoStop}%
\bibitem [{\citenamefont {Nakada}\ and\ \citenamefont
  {Sebe}(1996)}]{Nakada1996}%
  \BibitemOpen
  \bibfield  {author} {\bibinfo {author} {\bibfnamefont {H.}~\bibnamefont
  {Nakada}}\ and\ \bibinfo {author} {\bibfnamefont {T.}~\bibnamefont {Sebe}},\
  }\href@noop {} {\bibfield  {journal} {\bibinfo  {journal} {J. Phys. G: Nucl.
  Part. Phys.}\ }\textbf {\bibinfo {volume} {22}},\ \bibinfo {pages} {1349}
  (\bibinfo {year} {1996})}\BibitemShut {NoStop}%
\bibitem [{\citenamefont {Jokinen}\ \emph {et~al.}(1998)\citenamefont
  {Jokinen}, \citenamefont {\"{A}yst\"{o}}, \citenamefont {Dendooven},
  \citenamefont {Honkanen}, \citenamefont {Lipas}, \citenamefont
  {Per\"{a}j\"{a}rvi}, \citenamefont {Oinonen},\ and\ \citenamefont
  {Siiskonen}}]{Jokinen1998}%
  \BibitemOpen
  \bibfield  {author} {\bibinfo {author} {\bibfnamefont {A.}~\bibnamefont
  {Jokinen}}, \bibinfo {author} {\bibfnamefont {J.}~\bibnamefont
  {\"{A}yst\"{o}}}, \bibinfo {author} {\bibfnamefont {P.}~\bibnamefont
  {Dendooven}}, \bibinfo {author} {\bibfnamefont {A.}~\bibnamefont {Honkanen}},
  \bibinfo {author} {\bibfnamefont {P.}~\bibnamefont {Lipas}}, \bibinfo
  {author} {\bibfnamefont {K.}~\bibnamefont {Per\"{a}j\"{a}rvi}}, \bibinfo
  {author} {\bibfnamefont {M.}~\bibnamefont {Oinonen}}, \ and\ \bibinfo
  {author} {\bibfnamefont {T.}~\bibnamefont {Siiskonen}},\ }\href {\doibase
  10.1063/1.57369} {\bibfield  {journal} {\bibinfo  {journal} {AIP Conference
  Proceedings}\ }\textbf {\bibinfo {volume} {455}},\ \bibinfo {pages} {745}
  (\bibinfo {year} {1998})}\BibitemShut {NoStop}%
\bibitem [{\citenamefont {Sauvage-Letessier}\ \emph {et~al.}(1981)\citenamefont
  {Sauvage-Letessier}, \citenamefont {Quentin},\ and\ \citenamefont
  {Flocard}}]{Sauvage1981}%
  \BibitemOpen
  \bibfield  {author} {\bibinfo {author} {\bibfnamefont {J.}~\bibnamefont
  {Sauvage-Letessier}}, \bibinfo {author} {\bibfnamefont {P.}~\bibnamefont
  {Quentin}}, \ and\ \bibinfo {author} {\bibfnamefont {H.}~\bibnamefont
  {Flocard}},\ }\href {\doibase https://doi.org/10.1016/0375-9474(81)90074-9}
  {\bibfield  {journal} {\bibinfo  {journal} {Nuclear Physics A}\ }\textbf
  {\bibinfo {volume} {370}},\ \bibinfo {pages} {231} (\bibinfo {year}
  {1981})}\BibitemShut {NoStop}%
\bibitem [{\citenamefont {Bender}\ \emph {et~al.}(2003)\citenamefont {Bender},
  \citenamefont {Heenen},\ and\ \citenamefont {Reinhard}}]{Bender2003}%
  \BibitemOpen
  \bibfield  {author} {\bibinfo {author} {\bibfnamefont {M.}~\bibnamefont
  {Bender}}, \bibinfo {author} {\bibfnamefont {P.-H.}\ \bibnamefont {Heenen}},
  \ and\ \bibinfo {author} {\bibfnamefont {P.-G.}\ \bibnamefont {Reinhard}},\
  }\href {\doibase 10.1103/RevModPhys.75.121} {\bibfield  {journal} {\bibinfo
  {journal} {Rev. Mod. Phys.}\ }\textbf {\bibinfo {volume} {75}},\ \bibinfo
  {pages} {121} (\bibinfo {year} {2003})}\BibitemShut {NoStop}%
\bibitem [{\citenamefont {Audi}\ \emph {et~al.}(2017)\citenamefont {Audi},
  \citenamefont {Kondev}, \citenamefont {Wang}, \citenamefont {Huang},\ and\
  \citenamefont {Naimi}}]{NUBASE16}%
  \BibitemOpen
  \bibfield  {author} {\bibinfo {author} {\bibfnamefont {G.}~\bibnamefont
  {Audi}}, \bibinfo {author} {\bibfnamefont {F.~G.}\ \bibnamefont {Kondev}},
  \bibinfo {author} {\bibfnamefont {M.}~\bibnamefont {Wang}}, \bibinfo {author}
  {\bibfnamefont {W.}~\bibnamefont {Huang}}, \ and\ \bibinfo {author}
  {\bibfnamefont {S.}~\bibnamefont {Naimi}},\ }\href {\doibase
  10.1088/1674-1137/41/3/030001} {\bibfield  {journal} {\bibinfo  {journal}
  {Chinese Phys. C}\ }\textbf {\bibinfo {volume} {41}},\ \bibinfo {pages}
  {030001} (\bibinfo {year} {2017})}\BibitemShut {NoStop}%
\bibitem [{\citenamefont {Hilaire}\ and\ \citenamefont {Girod}(2007)}]{AMEDEE}%
  \BibitemOpen
  \bibfield  {author} {\bibinfo {author} {\bibfnamefont {S.}~\bibnamefont
  {Hilaire}}\ and\ \bibinfo {author} {\bibfnamefont {M.}~\bibnamefont
  {Girod}},\ }\href {\doibase https://doi.org/10.1051/ndata:07709} {\bibfield
  {journal} {\bibinfo  {journal} {International Conference on Nuclear Data for
  Science and Technology}\ ,\ \bibinfo {pages} {107}} (\bibinfo {year}
  {2007})}\BibitemShut {NoStop}%
\bibitem [{Mas()}]{MassExplorer}%
  \BibitemOpen
  \href@noop {} {\enquote {\bibinfo {title} {Mass explorer},}\ }\bibinfo {note}
  {{http://massexplorer.frib.msu.edu}}\BibitemShut {NoStop}%
\bibitem [{InP()}]{InPACS}%
  \BibitemOpen
  \href@noop {} {\enquote {\bibinfo {title} {Inpacs},}\ }\bibinfo {note}
  {Https://wwwnucl.ph.tsukuba.ac.jp/InPACS/}\BibitemShut {NoStop}%
\bibitem [{\citenamefont {Yoshida}(2015)}]{Yoshida2015}%
  \BibitemOpen
  \bibfield  {author} {\bibinfo {author} {\bibfnamefont {K.}~\bibnamefont
  {Yoshida}},\ }\href {\doibase 10.7566/JPSCP.6.020017} {\bibfield  {journal}
  {\bibinfo  {journal} {JPS. Conf. Proc.}\ }\textbf {\bibinfo {volume} {6}},\
  \bibinfo {pages} {020017} (\bibinfo {year} {2015})},\ \Eprint
  {http://arxiv.org/abs/https://journals.jps.jp/doi/pdf/10.7566/JPSCP.6.020017}
  {https://journals.jps.jp/doi/pdf/10.7566/JPSCP.6.020017} \BibitemShut
  {NoStop}%
\end{thebibliography}%

\end{document}